\documentclass[10pt,twocolumn]{article} 
\pdfoutput=1
\usepackage{simpleConference}
\usepackage{times,natbib}
\usepackage{graphicx}
\usepackage{amssymb}
\usepackage{url,hyperref}
\usepackage{cite}
\usepackage{amsmath}
\usepackage{tensor}
\usepackage{booktabs}

\renewcommand{\vec}[1]{\boldsymbol{\mathrm{#1}}}

\begin{document}

\title{Fast and Accurate Computation of Vertical Modes}

\author{Jeffrey J. Early, NorthWest Research Associates, USA \\
M. Pascale Lelong, NorthWest Research Associates, USA \\
K. Shafer Smith, New York University, USA}

\maketitle
\thispagestyle{empty}

\abstract{The vertical modes of linearized equations of motion are widely used by the oceanographic community in numerous theoretical and observational contexts. However, the standard approach for solving the generalized eigenvalue problem using second-order finite difference matrices produces $O(1)$ errors for all but the few lowest modes, and increasing resolution quickly becomes too slow as the computational complexity of eigenvalue algorithms increase as $O(n^3)$. Existing methods are therefore inadequate for computing a full spectrum of internal waves, such as needed for initializing a numerical model with a full internal wave spectrum. Here we show that rewriting the eigenvalue problem in stretched coordinates and projecting onto Chebyshev polynomials results in substantially more accurate modes than finite-differencing at a fraction of the computational cost. We also compute the surface quasigeostrophic modes using the same methods. All spectral and finite difference algorithms are made available in a suite of Matlab classes that have been validated against known analytical solutions in constant and exponential stratification.}
\\ \\
\noindent This work has not yet been peer-reviewed and is provided by the contributing author(s) as a means to ensure timely dissemination of scholarly and technical work on a noncommercial basis. Copyright and all rights therein are maintained by the author(s) or by other copyright owners. It is understood that all persons copying this information will adhere to the terms and constraints invoked by each author's copyright. This work may not be reposted without explicit permission of the copyright owner.

%
\section{Introduction}
%

Vertical modes arise as part of the separable solution to both the internal wave problem and quasigeostrophic theory. The eigenvalue problem (EVP) is treated in many introductory physical oceanography textbooks, e.g., \citet{gill1982-book,cushman2011-book}, and the resulting vertical modes describe the vertical structure of the linear solutions for a given density profile. While there are an infinite number of bases that can be used to represent ocean currents and density anomalies satisfying certain boundary conditions, the vertical modes correspond to $O(1)$ dynamical solutions of the equations of motion, and are therefore both diagnostic and prognostic. For this reason, vertical modes are the standard basis with which to represent the vertical structure of ocean currents, and it would be hard to overstate their usefulness for describing and modeling the ocean.

There are two primary uses for the vertical modes: (1) projecting a given flow field onto the vertical modes to determine its spectrum (a forward transformation), or (2) creating a dynamically consistent linear flow field from a given spectrum (an inverse transformation). For example, projection onto the vertical modes was used to construct the Garrett-Munk internal wave spectrum \citep{munk1981-book,polzin2011-rg}, as well as to describe the vertical structure of balanced flow in the ocean \citep{wunsch1997-jpo,wortham2014-jpo}.

This study was motivated by two situations where current techniques for computing vertical modes were found to have significant errors for reasonable computation times. In the first situation a numerical model needed to be initialized with a full spectrum of internal wave---a task which requires solving an eigenvalue problem at each resolved wavenumber in the model. In the second situation, we sought to compare an observed horizontal velocity spectrum of internal waves to the Garrett-Munk spectrum near a very strong pycnocline. This requires computing vertical modes at high frequencies---which requires appropriately resolving the mode variability near the pycnocline.

There are a number of sources of error that arise in performing either the forward or inverse transformation. These include the following:
\begin{enumerate}
\item a poorly defined mean density function;
\item measurement noise and uncertainty in the density function and, in the case of a forward transform, the dynamical variables;
\item aliasing error, due to the location of the grid points of the dynamical variables (relevant only for forward transform);
\item interpolation error, due to the location of (and lack of) data points specifying the density function;
\item numerical truncation error of the modes generated from the vertical eigenvalue problem.
\end{enumerate}
Whether performing a forward or inverse transformation, these sources of error will compound in some fashion to create error in either the resulting flow field or the resulting spectrum.

The first source of error, a poorly defined mean density function, can result from lack of data, or often just uncertainty in what qualifies as a mean (e.g., questions of what time and length scales to average over, or even whether averaging is the correct approach for a nonlinear system). Measurement noise and uncertainty in data is a topic unto itself, so the methods here proceed without concern for measurement noise when possible. The aliasing error arises when the grid points are placed such that higher modes project onto the lower modes. This source of error is relatively easily controlled by computing the condition number of the projection matrix, which provides a fairly precise cutoff for the number of resolvable modes. On the rare occasion that a density function can be specified analytically, the interpolation error can be minimized to numerical precision. However, in the usual case where the density is given on some grid with uneven spacing, the density function must be interpolated in between those grid points. Our work suggests density interpolation error does not dominate the error for most cases. This manuscript is therefore largely concerned with the final source of errors---arising from the numerical representation of the modes in the vertical eigenvalue problem.

The standard method for solving the vertical eigenvalue problem is to discretize the problem and construct the matrices using second-order finite difference matrices, e.g., \citet{cushman2011-book}. However, this approach produces unacceptable errors for all but the very lowest modes in the simplest stratifications.  This is problematic because numerical algorithms for solving eigenvalue problems scale as $O(n^3)$, where $n$ is the number of discretization points in the vertical, so small increases in resolution come at a large computational cost. Instead of using finite difference methods, \citet{kelly2016-jpo} solves the hydrostatic form of the eigenvalue problem spectrally using Galerkin's method, at a fraction of the computational cost and with much higher accuracy. \citet{dunphy2009-thesis} uses a Chebyshev collocation method to solve the non-hydrostatic case with fixed frequency. Here, we extend the ideas of \citet{kelly2016-jpo} and \citet{dunphy2009-thesis} to the more general non-hydrostatic cases where the EVP must be solved for each frequency in the spectrum or each resolved wavenumber in a numerical model. Our approach solves the eigenvalue problem spectrally with Chebyshev polynomials to produce high quality vertical modes, even with relatively small $n$. The same techniques are then applied to solve for surface quasigeostrophic modes, used to describe the effect of density anomalies at the ocean boundaries.

This paper begins with a derivation of the two most relevant forms of the vertical eigenvalue problem that arise from the linearized equations of motion in section \ref{sec:linear_iw}. This provides the necessary context for orthogonality relations that form the basis of the normalization of the vertical modes, which in turn shows limitations of using certain vertical modes as a basis. Section \ref{sec:prob_with_fd} demonstrates some of the problems associated with finite differencing, while section \ref{sec:z_coord} shows how these can be remedied using spectral methods. Details of the numerical implementation are described in section \ref{sec:implementation} and some of the other sources of error are examined in section \ref{sec:error}. Section \ref{sec:discussion} discusses some best practices and potential pitfalls. Finally, the appendices include the exact analytical solutions for constant and exponential stratification that are employed for unit testing, the WKB (Wentzel-Kramers-Brillouin) approximated solution used for the adaptive grid method, and the class inheritance tree of the publicly available Matlab implementation of these methods.

%
\section{Background}
\label{sec:linear_iw}
%

The linearized equations of motion for the fluid velocity $u(x,y,z,t)$, $v(x,y,z,t)$, $w(x,y,z,t)$, on the $f$-plane are
\begin{align}
\label{x-momentum}
\partial_t u - f_0 v =& - \frac{1}{\rho_0} \partial_x p \\ \label{y-momentum}
\partial_t v + f_0 u =& - \frac{1}{\rho_0} \partial_y p \\ \label {z-momentum}
\partial_t w =& - \frac{1}{\rho_0} \partial_z p - g \frac{\rho}{\rho_0} \\ \label{continuity}
\partial_x u + \partial_y v + \partial_z w =& 0 \\ \label{thermodynamic}
\partial_t \rho + w \partial_z \bar{\rho} =& 0
\end{align}
where $p(x,y,z,t)$ and $\rho(x,y,z,t)$ are the perturbation pressure and density, respectively. These are defined such that the total pressure $p_{\textrm{tot}}(x,y,z,t) = p(x,y,z,t) + p_0(z)$ and the total density $\rho_{\textrm{tot}}(x,y,z,t) = \rho_0 + \bar{\rho}(z) + \rho(x,y,z,t) $ where $\partial_z p_0(z) = -g \bar{\rho}(z)$. All variables in the equations of motion are functions of $x$, $y$, $z$ and $t$, except for $\bar{\rho}$ which is strictly a function of $z$. The operator $\partial_z$ is understood to reduce to $\frac{d}{dz}$ when applied to univariate functions. We use the usual definition of the buoyancy frequency $N^2(z) \equiv -\frac{g}{\rho_0} \partial_z \bar{\rho}$.


There are three linearly independent solutions to equations \ref{x-momentum}-\ref{thermodynamic}, assuming periodic horizontal boundary conditions and a flat bottom: two wave solutions and the geostrophic solution.

\subsection{Wave solutions}
\label{sec:wave_soln}
The positive frequency wave solution is given by,
\begin{equation}
\label{positive_wave_solution}
\left[\begin{array}{c} p_+ \\ u_+ \\ v_+ \\ w_+ \\ \rho_+  \end{array}\right] =
A \left[\begin{array}{c}
- \rho_0 g \frac{K h}{\omega} \cos \theta_+ F(z)\\
	\left(\cos \alpha \cos \theta_+ + \frac{f_0}{\omega} \sin \alpha \sin \theta_+\right)F(z) \\
	\left(\sin \alpha \cos \theta_+ - \frac{f_0}{\omega} \cos \alpha \sin \theta_+\right)F(z) \\
    K h \sin \theta_+ G(z) \\
	\frac{d \bar{\rho}}{d z} \frac{K h}{\omega} \cos \theta_+G(z)
 \end{array}\right]
\end{equation}
where the functions $F(z)$ and $G(z)$ are the eigenfunctions with corresponding eigenvalue $h$, to be discussed in detail below. The frequency is determined through the dispersion relation,
\begin{equation}
\omega = \sqrt{g h (k^2 + l^2) + f_0^2}
\end{equation}
and the negative rotating wave solution is found by flipping the sign on the frequency, $\omega \mapsto -\omega$. In this notation the phase angle of the wave is given by $\alpha = \tan^{-1} \left(\frac{l}{k}\right)$ and $K = \sqrt{k^2 + l^2}$ is the total horizontal wavenumber. The horizontal phase is given by $\theta_\pm=k x + l y \pm \omega t$. The eigenvalue $h$ is referred to as the equivalent depth and is related to the wave group velocity, $c_g = \sqrt{gh}$. The value $h$  can be replaced in favor of eigenfrequency $\omega$ using the dispersion relation, but here we include both to avoid singularities at $K=0$ and for notational compactness.  Applying this solution to equations \ref{x-momentum}-\ref{thermodynamic} leads to two coupled equations for the vertical structure functions,
\begin{equation}
(N^2-\omega^2)G = -g \partial_z F \quad\text{and}\quad
F = h \partial_z G,
\end{equation}
which can be combined into various second-order eigenvalue problems (see section \ref{sec:vertical_evp}).


\subsection{Geostrophic solution}
\label{sec:geostrophic_soln}
The geostrophic solution is given by,
\begin{equation}
\label{geostrophic_solution}
\left[\begin{array}{c} p_g \\ u_g \\ v_g \\ w_g \\  \rho_g  \end{array}\right] =
B \left[\begin{array}{c}
 \rho_0 g \cos \theta_0 F(z) \\
	\frac{g}{f_0} l \sin \theta_0 F(z) \\
	-\frac{g}{f_0} k \sin \theta_0 F(z) \\
    0 \\
	-\frac{d \bar{\rho}}{d z} \cos \theta_0 G(z)
 \end{array}\right]
\end{equation}
where $\theta_0=k x + l y$. The solution can also be written entirely in terms of streamfunction $\psi(x,y,z) = \frac{g}{f_0} \cos \theta_0 F(z)$ where $(u_g,v_g,\rho_g)=(-\partial_y \psi, \partial_x \psi, -\rho_0 f_0 \partial_z \psi/g)$. Satisfying the vertical momentum equation requires that $N^2 G = -g \partial_z F$ but, unlike the wave solutions, geostrophic solutions already satisfy continuity. For a given wavenumber $(k,l)$ the geostrophic solution is entirely specified by a vertical profile of any one of the variables, from which all others are immediately deduced. For example, $\hat{\rho}(k,l,z)$ determines $G(z)$, from which $F(z)$ is determined by integration---the thermal wind balance. There is no preferred basis for the geostrophic solution.

Although the scalings that lead to equations \ref{x-momentum}-\ref{thermodynamic} result in the linear geostrophic solution where $w_g=0$, near-geostrophic theories with a different choice of scalings, such as quasi-geostrophy \citep{pedlosky1987-book}, have nonzero vertical velocities $w_g \neq 0$ and therefore require full continuity, i.e., that $F = h \partial_z G$. As with the wave solution, this requirement combined with $N^2 G = -g \partial_z F$ results in an eigenvalue problem, detailed in section \ref{sec:vertical_evp}.

An eigenbasis constructed with the rigid lid boundary condition ($w(0)=G(0)=0$) precludes non-zero density anomalies at the surface. One workaround to this limitation is to further decompose the geostrophic solution into three parts: two parts resulting from the density anomaly at the boundaries and one part from the remaining density anomaly in the interior. Following \citet{lapeyre2006-jpo}, the idea is then to let
\begin{equation}
\psi = \psi^{\textrm{int}} + \psi^{\textrm{sur}} + \psi^{\textrm{bot}}
\end{equation}
where both the surface and bottom components of the flow are required to have no potential vorticity in the interior,
\begin{equation}
\label{sqg_eqn}
\nabla^2 \psi^{\textrm{sur/bot}} + \partial_z \left( \frac{f_0^2}{N^2} \partial_z \psi^{\textrm{sur/bot}} \right) = 0,
\end{equation}
but account for the density anomaly at the boundaries, e.g.,
\begin{equation}
f_0 \partial_z \psi^{\textrm{sur}} = -\frac{g}{\rho_0} \rho \bigr\rvert_{z=0}, \quad f_0 \partial_z \psi^{\textrm{bot}} = 0 \bigr\rvert_{z=-D}.
\end{equation}
The resulting modes can be solved for a given wavenumber and are referred to as the surface quasi-geostrophic (SQG) modes. This methodology has been used to construct the interior velocity field from sea-surface height and temperature data \citet{wang2013-jpo}.

\citet{smith2013-jpo} formulate a new eigenvalue problem that results in modes capable of capturing surface density anomalies for quasigeostrophic flows. Taking equation 9 from \citet{smith2013-jpo} and writing it in the present notation we have,
\begin{equation}
\label{vertical-eigenvalue-SV}
\tag{SV EVP}
\partial_{zz}G_j - \frac{K^2 N^2}{f_0^2} G_j = -\frac{N^2}{g h_j }G_j
\end{equation}
with surface boundary condition $G(0)=\frac{D}{\alpha}\partial_z G$ where $\alpha$ is a tunable parameter. Importantly, these modes remain orthogonal, unlike the combined set of SQG modes and interior modes described above.

An alternative to both the SQG and the \citet{smith2013-jpo} approaches is to use the internal wave eigenbasis constructed with the free surface boundary condition (detailed below) which also results in orthogonal modes capable of capturing nonzero density anomaly at the ocean surface.

\subsection{Vertical eigenvalue problem}
\label{sec:vertical_evp}
The vertical eigenvalue problem is formed using the two coupled equations from section \ref{sec:wave_soln} for the vertical structure functions, $(N^2-\omega^2)G = -g \partial_z F$ (vertical momentum) and $F = h \partial_z G$ (continuity). In combination with the dispersion relation, one of the eigenfunctions can be eliminated, resulting in various single second-order eigenvalue problems for the vertical structure functions $F$ or $G$. The two most practical eigenvalue problems to solve are for $G(z)$ with constant $\omega$,
\begin{equation}
\label{vertical-eigenvalue-G-with-omega}
\tag{$\omega$-constant EVP}
\partial_{zz} G_j = -\frac{N^2-\omega^2}{g h_j }G_j
\end{equation}
and for $G(z)$ with constant $K$,
\begin{equation}
\label{vertical-eigenvalue-G-with-K}
\tag{$K$-constant EVP}
\partial_{zz}G_j - K^2 G_j = -\frac{N^2 - f_0^2}{g h_j }G_j.
\end{equation}
The near geostrophic eigenvalue problem is found by combining the vertical momentum condition $N^2G = -g \partial_z F$ with the continuity condition $F = h \partial_z G$, which can be treated as the \ref{vertical-eigenvalue-G-with-omega} with $\omega=0$. Note that this is equivalent to making the hydrostatic approximation \citep{kelly2016-jpo}, and removes all dependence on frequency $\omega$ and wavenumber $K$.

For the cases considered here we take the lower boundary condition at $z=-D$ to be either free-slip, where $w(-D)=0$, or no-slip, where $u(-D)=0$. These correspond to $G(-D)=0$ and $F(-D)=0$, respectively. These conditions can be seen as the limiting cases of sloped bottom topography \citep{lacasce2017-grl}. The surface boundary condition at $z=0$ is taken to be either a rigid lid with $w(0)=0$, $G(0)=0$, or a free surface approximated as $p(x,y,0)=\rho_0 g \eta(x,y,0)$ where $\eta\equiv-(\partial_z \bar{\rho})^{-1} \rho$ is the linear approximation of the isopycnal displacement. In terms of the vertical modes, the free surface boundary condition becomes $ h \partial_z G(0) = G(0)$. Finally, there are many cases where the density profile does not extend to the full depth of the ocean and no boundary conditions (beyond the EVP itself) should be added.

Solving either EVP yields a set of eigenvalues $h_j$ that can be ordered such that $h_1>h_2>h_3>..>h_n$, each with corresponding eigenfunction $G_j$. This means that solving \ref{vertical-eigenvalue-G-with-omega} results in wave solutions with the same frequency $\omega$, but different wavenumbers $K_j$, and similarly solving \ref{vertical-eigenvalue-G-with-K} results in wave solutions with the same wavenumber $K$ and different frequencies $\omega_j$. Although we do not implement this numerically, note that rearranging the \ref{vertical-eigenvalue-G-with-K} poses the EVP for fixed group velocity, $gh$, with eigenvalue $K^2_j$.

The equations for \ref{vertical-eigenvalue-G-with-omega} and \ref{vertical-eigenvalue-G-with-K} are both Sturm-Liouville problems and share the property that their eigenmodes are orthogonal. Following the procedure in \citet{kelly2016-jpo}, the eigenmodes found with the \ref{vertical-eigenvalue-G-with-omega} satisfy
\begin{equation}
\label{omega_const_ortho}
G_i(0)G_j(0)+\frac{1}{g} \int_{-D}^{0} (N^2(z)-\omega^2) G_i G_j \, dz = \beta \delta_{ij} 
\end{equation}
and
\begin{equation}
\int_{-D}^0   F_i F_j  \, dz = \beta h_i \delta_{ij} 
\end{equation}
while the eigenmodes found with \ref{vertical-eigenvalue-G-with-K} satisfy,
\begin{equation}
\label{k_const_ortho}
G_i(0)G_j(0)+\frac{1}{g} \int_{-D}^{0} (N^2(z)-f_0^2) G_i G_j \, dz = \gamma \delta_{ij}
\end{equation}
and
\begin{equation}
\int_{-D}^0  \left( F_i F_j + h_i h_j K^2 G_i G_j \right) \, dz = \gamma h_i \delta_{ij} 
\end{equation}
where $\beta$ and $\gamma$ are unspecified constants that depend on the chosen normalization, as discussed below. It is important to note that these orthogonality conditions only apply for a particular choice of $\omega$ or $K$. For example, an eigenmode $G_j(z,k_1)$ found using $K=k_1$ is not orthogonal to an eigenmode $G_j(z,k_2)$ found using $K=k_2$ if $k_1 \neq k_2$.

The most significant difference between the two EVPs is that eigenmodes from the \ref{vertical-eigenvalue-G-with-K} often form a complete basis set for typical ocean stratification profiles, while the eigenmodes from the \ref{vertical-eigenvalue-G-with-omega} do not. The \ref{vertical-eigenvalue-G-with-K} is a \emph{regular} Sturm-Liouville problem when the weighting function $w_K(z) \equiv N^2 - f_0^2 >0$ for all $z$, a condition typically met in the ocean. We note that although it is fair to say that stratification with $N>f_0$ is typical of the world oceans, after examining 30,000 CTD profiles \citet{kunze2017-jpo} finds that 10\% of the data have $N<2f$ and a full 30\% of the data suggest $N<2f$ within 380 meters of the bottom. In contrast, the weighting function $w_\omega(z) \equiv N^2-\omega^2$ in the \ref{vertical-eigenvalue-G-with-omega} switches sign at turning points $z_T$, where $N(z_T)=\omega$. Consequently, the norm of an arbitrary function defined on the domain $[-D,0]$ and satisfying the boundary conditions is not guaranteed to be positive using the norm implied by equation \ref{omega_const_ortho}, a necessary condition for completeness. Intuitively this can be seen in figure \ref{PycnoclineMode}, which shows that the high frequency modes have no variance beyond the turning points, and are therefore incapable of representing arbitrary functions on the domain.

\subsection{Normalization}
\label{sec:normalization}

The amplitude of each vertical mode can be scaled by an arbitrary constant, so one must choose a normalization appropriate for the problem being considered. The four most common scenarios are setting the total energy, a horizontal velocity ($U$),  a vertical velocity ($W$), and the sea surface height (SSH) of a given wave.

To set the total energy of the internal wave solution in equation \ref{positive_wave_solution}, we use the modes from the \ref{vertical-eigenvalue-G-with-K} and therefore use the norm implied by equation \ref{k_const_ortho},
\begin{equation}
\label{k_const_norm}
G_i^2(0)+\frac{1}{g} \int_{-D}^{0} (N^2(z)-f_0^2) G_i^2 \, dz = 1 \tag{$K$-constant norm}
\end{equation}
where we have chosen $\gamma=1$ as the normalization constant in order to keep the vertical modes unitless.\footnote{Another reasonable choice is to take $\gamma = N_0^2 D/g$, but using $\gamma=1$ keeps the norm universal, rather than problem specific.} Taking the total energy of the wave,
\begin{equation}
 E(x,y,z,t) = \frac{1}{2}\left(u^2 + v^2 + w^2 + \frac{g^2}{\rho_0^2}\frac{\rho^2}{N^2}\right)
\end{equation}
and then depth integrating and averaging over space and time produces a wave with energy $P^2/2$ if we set the coefficient $A=P/\sqrt{h_j - (\omega^2 - f_0^2)G^2(0)/\omega^2}$ in equation \ref{positive_wave_solution}.

Setting the maximum initial eastward velocity to $U$ can be accomplished by imposing $\max{F_j}=1$ and $A=U$. The maximum vertical velocity $W$ is set using the norm $\max{G_j}=1$ where $A=W/(Kh_j)$, but is clearly singular for inertial waves which have no vertical velocity. The sea surface height is set using the pressure at the surface with $F(0)=1$ and $A=\textrm{SSH} \cdot \frac{\omega}{K h}$.

To set the total energy of the interior geostrophic solution in equation \ref{geostrophic_solution}, we assume the solution uses the typical geostrophic modes from equation \ref{vertical-eigenvalue-G-with-omega} with $\omega=0$ and therefore use the norm implied by equation \ref{omega_const_ortho},
\begin{equation}
\label{omega_const_norm}
\frac{1}{D} \int_{-D}^{0} F_i^2 \, dz = 1
\tag{$\omega$-constant norm}
\end{equation}
where we have taken $\beta = D/h_i$. This produces a mode with energy $P^2$ if we let,
\begin{equation}
B^2 = \frac{4 P^2 f^2 h}{g D} \frac{1}{ghK^2 + f^2(1+h G^2(0)/D)}.
\end{equation}
Setting the maximum eastward velocity requires $B = U \frac{f_0}{g l}$ using $\max{F_j}=1$. The sea-surface height is set using the pressure at the surface by setting $F(0)=1$ and $B=\textrm{SSH}$.

%
\section{The problem with finite differencing}
\label{sec:prob_with_fd}
%

Computing the lowest vertical modes with finite differencing methods can be relatively fast and accurate when considering a single wavenumber or frequency. Although one can encounter problems with the higher modes, this can usually be ameliorated by increasing resolution. The primary issues with finite differencing arise when needing to compute many modes across a broad range of frequencies and wavenumbers---the two scenarios that motivated the present study. To compute a complete internal wave frequency spectrum requires solving the \ref{vertical-eigenvalue-G-with-omega} at each resolved frequency between the Coriolis frequency and the maximum buoyancy frequency, roughly $O(10^2)$ EVPs. This is especially challenging near the buoyancy frequency, where all oscillations occur in the narrow region where $N(z)<\omega$. On the other hand, initializing a numerical model with an internal wave spectrum involves solving the \ref{vertical-eigenvalue-G-with-K} for each resolved wavenumber in the model, which easily requires $O(10^4)$ computations or more.

\begin{figure}[t]
  \centerline{\includegraphics[width=19pc,angle=0]{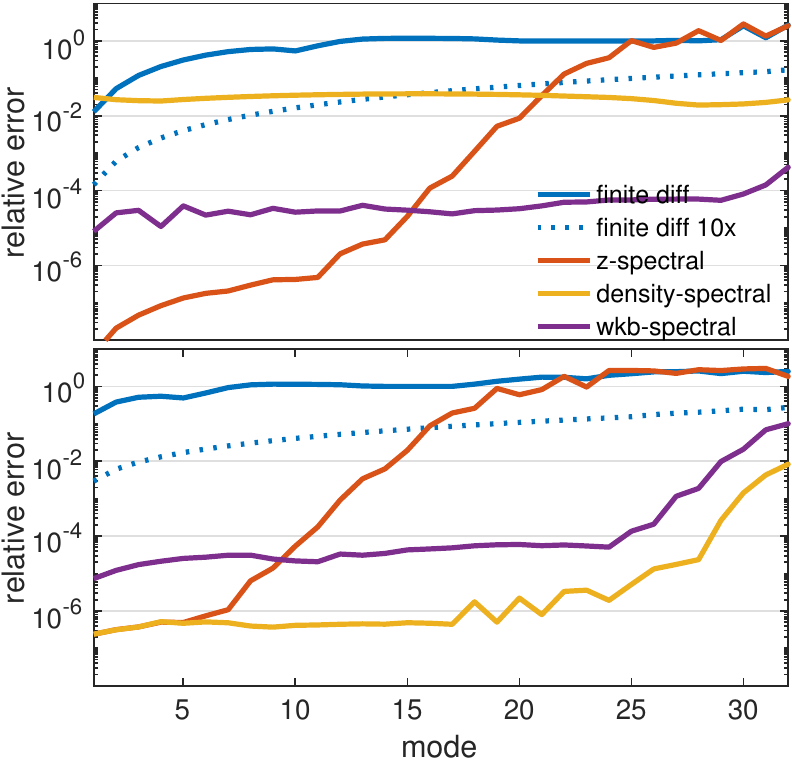}}
  \caption{Relative error as a function of vertical mode number using 64 evenly spaced grid points in exponential stratification $N(z)=N_0 e^{z/b}$ where $N_0=3$ cph and $b=1300$ m at latitude 33$^\circ$ in a 5000 m deep ocean for $K=0.0$ (top panel) and $K=\frac{2\pi}{500\textrm{ m}}$ (bottom panel). Shown are two 2nd-order finite differencing methods with (1) 64 grid points (blue), (2) 631 grid points (blue dotted), and three spectral methods using Chebyshev polynomials with coordinates in (3) depth (red), (4) WKB scaled (purple), and (5) density (orange).}
  \label{InternalModeError}
\end{figure}

A prerequisite to initializing a numerical model with a given internal waves spectrum, is that the modes must be computed for each unique horizontal wavenumber $K$ resolved by the model using the \ref{vertical-eigenvalue-G-with-K}. If the numerical model has $(N_x,N_y)$ horizontal grid points, approximately $N_x N_y/2$ unique eigenvalue problems must be solved (up to another factor of 2 can be eliminated with isotropic horizontal resolution). Unfortunately, eigenvalue algorithms scale as $O(n^3)$ for $n$ by $n$ matrices. This means that initialization of an internal wave spectrum scales as $O(N_x N_y N_z^3)$ and thus, with any reasonable vertical resolution, this will quickly become a rate limiting step to a model run. In practical terms, the computation time of these eigenvalue problems takes approximately 1s, 10s, and 100s, for $N_z$ of 512, 1024, and 2048, respectively, on consumer hardware from 2015.

The problem is further exacerbated by the poor performance of finite difference methods. To demonstrate, we compare different numerical methods against an analytical solution. Consider an exponential density profile---the canonical deep ocean stratification profile which has known analytical solutions for both the non-hydrostatic internal modes \citep{garrett1972-gfd}, as well as the SQG modes \citep{lacasce2012-jpo}. Using a numerical method to solve the \ref{vertical-eigenvalue-G-with-K}, we can compute the relative error of the numerical solution to the analytical solution. We define the relative error as
\begin{equation}
\textrm{rel. err.} = \frac{\max(|f_i - f(z_i)|)}{\max(f(z_i))}
\end{equation}
where $f(z_i)$ is the true solution evaluated at the grid points $z_i$, and $f_i$ is the numerical approximation. Figure \ref{InternalModeError} shows the maximum relative error of the eigenmodes $F$, $G$ and eigenvalue $h$ found by solving the \ref{vertical-eigenvalue-G-with-K} using 2nd order finite difference methods for standard exponential stratification with 64 vertical grid points (blue). Details of the numerical implementation of the analytical solutions are given in \ref{sec:im_exponential_strat}.

Reasonable error magnitudes are $O(10^{-2})$ (see \ref{sec:mode_error} for justification), however, the top panel of figure \ref{InternalModeError} shows that no modes computed with finite differencing satisfy this condition. The situation is even worse for the $K=\frac{2\pi}{500\textrm{ m}}$ case shown in the bottom panel, where even the lowest mode has an $O(0.1)$ error. The blue dotted line in figure \ref{InternalModeError} shows that increasing the resolution tenfold for 2nd-order finite differencing decreases the error by a factor of 100, as would be expected. However, this comes at 1000 times the computational cost, and still barely produces any usable modes.

\begin{figure}[t]
  \centerline{\includegraphics[width=19pc,angle=0]{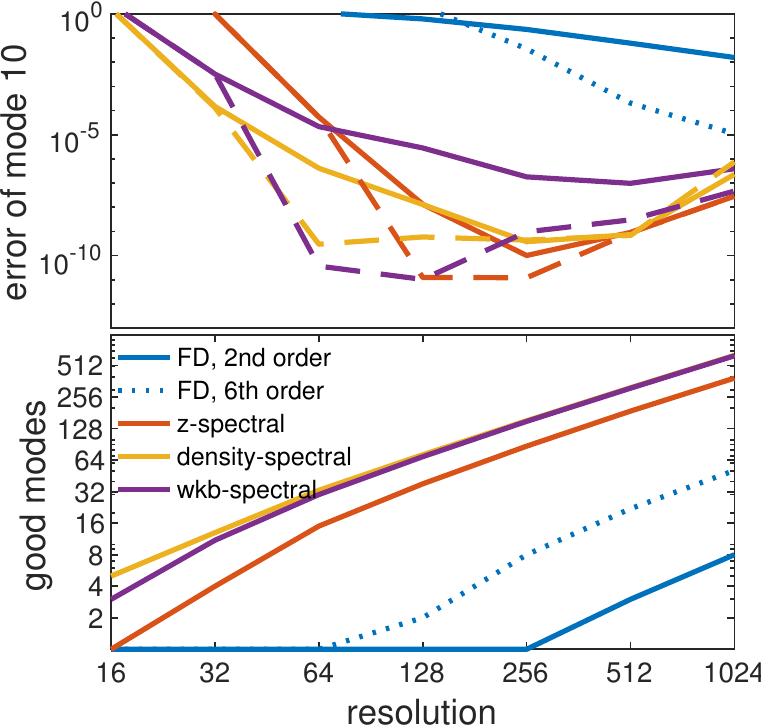}}
  \caption{The top panel shows relative error as a function of resolution for the 10th mode in exponential stratification with $K=\frac{2\pi}{500\textrm{ m}}$. The density function is specified on an evenly spaced grid (solid lines) or passed directly as an analytical function (dashed lines). The bottom panel shows the number of usable modes as a function of resolution, defined as the number of modes with truncation errors less than $10^{-2}$. The convergence rate of the 2nd order and 6th order finite difference methods are found to be $(\Delta z)^{2.0}$ and $(\Delta z)^{5.8}$, respectively. }
  \label{TruncationErrorK2pi500Mode10}
\end{figure}

The accuracy of finite differencing can be increased by going to higher orders \citep{fornberg1998-siam}, since the truncation error at order $s$ scales as $(\Delta z)^s$. The truncation error of the 10th mode in exponential stratification is shown in the top panel of figure \ref{TruncationErrorK2pi500Mode10} where the blue (solid, dotted) line shows the (2nd, 6th) order finite difference method converging at its predicted rate. The bottom panel shows that even with 1024 grid points, there are only 8 usable modes for the 2nd order finite differencing method, while 6th order gives up to 50 modes. However, while increasing the order of the method does provide some gains in accuracy, the most efficient way to proceed is simply to use spectral methods, which promise exponentially decreasing truncation error, rather than the polynomial truncation errors offered by finite differencing. When using an analytical density function (dashed lines, figure \ref{TruncationErrorK2pi500Mode10}) rather than gridded data, there is no interpolation error and the spectral methods truncation errors reach a noise floor somewhere between 64 and 128 grid points. Furthermore, the number of usable modes is an order of magnitude higher than even the 6th order finite difference method. In practical terms, the 2nd order finite difference method is producing about 10 good modes in 100 seconds, while the spectral methods are producing about 100 good modes in 1 second. The increase in truncation error at higher resolution is likely due to increasingly compounded errors of the eigenvalue solvers.

%
\section{Chebyshev polynomials}
\label{sec:z_coord}
%

Written in matrix form the \ref{vertical-eigenvalue-G-with-K} is,
\begin{equation}\label{EVPforAB}
\mathsf{A} \vec{v} = \frac{1}{h} \mathsf{B} \vec{v}
\end{equation}
where $\vec{v}$ is the vector representation of the normal mode $G$ at grid points in $z$, $\mathsf{A}=\partial_{zz} - K^2$ and $\mathsf{B} = (f_0^2 - N^2)/g$. For finite differencing, $\mathsf{A} = \mathsf{D}_{zz} - K^2 \mathsf{I}$, where $\mathsf{D}_{zz}$ is the $N_z \times N_z$ differentiation matrix and $\mathsf{I}$ is the $N_z$ dimensional identity matrix. To use Chebyshev polynomials, we project vector $\vec{v}$ onto a Chebyshev basis using $\hat{\vec{v}} = \mathsf{T}^{-1}\vec{v}$ where $\mathsf{T}$ is the matrix that transforms a vector from a Chebyshev basis to the coordinate basis. In a practical sense, the columns of $\mathsf{T}$ are the Chebyshev polynomials. Then the eigenvalue problem becomes,
\begin{equation}
\mathsf{A} \mathsf{T} \mathsf{T}^{-1}\vec{v} = \frac{1}{h} \mathsf{B} \mathsf{T} \mathsf{T}^{-1}\vec{v}
\end{equation}
or simply,
\begin{equation}\label{EVPforT}
(\partial_{zz} \mathsf{T} - K^2 \mathsf{T}) \hat{\vec{v}} = \frac{1}{h} \left(\frac{f_0^2 -N^2(z)}{g} \right) \mathsf{T} \hat{\vec{v}}.
\end{equation}
The vector $\hat{\vec{v}}$ contains the coefficients needed to reconstruct eigenfunctions and $\partial_{zz} \mathsf{T}$ are the second derivatives of the Chebyshev poynomials.

The optimal choice of grid for Chebyshev polynomials is a Gauss-Lobatto grid \citep{boyd2001-book,canuto2006-book}, e.g. equation \ref{lobatto} below, and thus the eigenmatrices and eignfunctions are always created on a Gauss-Lobatto grid for any chosen coordinate. Because the basis functions are continuous functions of $z$, the resulting vertical modes can be interpolated onto any grid at any resolution by evaluating the functions at the points of interest.

It is, however, rarely the case that density is given as an analytical function, or that observations are made on a Gauss-Lobatto grid, which means that typically the density needs to be interpolated on the appropriate grid. Interpolation is performed using B-splines implemented with the numerical framework described in \citet{early2019-arxiv}. The advantage to using B-splines to represent gridded density data is that it is easy to accommodate arbitrary grids. Despite being a low order method, this is generally not a limitation (see section \ref{sec:error}).  In the cases shown in figure \ref{InternalModeError}, the algorithms are given the density ($\rho$) on a uniform grid in $z$ of 64 points and the resulting modes are returned on the same grid (except where noted for the high resolution finite differencing case). This is, of course, suboptimal for the spectral cases which use a Gauss-Lobatto grid on various coordinates to compute the eigenvalue problem. When given an analytical function for density, these methods perform even better, as can be seen in figure \ref{TruncationErrorK2pi500Mode10}.

Despite the potential limitations imposed by interpolating the density with B-splines onto a Gauss-Lobatto grid, the red line in figure \ref{InternalModeError} shows that the Chebyshev method performs extremely well, even while interpolating from an evenly-spaced grid, and outputting to the same grid. The first 20 and 14 modes have error less than $O(10^{-2})$ for the $K=0$ and $K=2\pi/500 \textrm{ m}^{-1}$ cases, respectively. However, at higher horizontal resolution (larger wavenumbers $K$), even the spectral method's errors grow large, because the points at which the functions are evaluated do not sufficiently capture the oscillations of the modes. This can be remedied by using a stretched coordinate, $s$.

\subsection{Stretched coordinates}
\label{sec:stretched_coord}

In order to find an independent coordinate better suited to capturing the structure of the eigenmodes, we rewrite the eigenvalue problems in terms of a generic coordinate $s$ and then consider two concrete examples. Taking $z$ to be a function of $s$ and applying the chain rule leads to
\begin{equation}
\partial_z = (\partial_s z)^{-1} \partial_s
\end{equation}
and
\begin{equation}
\partial_{zz} = - (\partial_{ss}z) (\partial_s z)^{-3} \partial_s + (\partial_s z)^{-2} \partial_{ss}.
\end{equation}
For example, the \ref{vertical-eigenvalue-G-with-K} becomes,
\begin{equation}
\label{evp_stretched}
\left( - (\partial_{ss}z) (\partial_s z)^{-3} \partial_s + (\partial_s z)^{-2} \partial_{ss} \right) G - K^2 G = -\frac{N^2 - f_0^2}{gh}G
\end{equation}
where now $F = h (\partial_s z)^{-1} \partial_s G$. The free surface boundary condition in these coordinates becomes $(\partial_s z)^{-1} \partial_s G = G/h$, while the normalization conditions are now,
\begin{equation}
\label{k_const_norm_stretched}
 \frac{1}{g} \int_{s(-D)}^{s(0)} (N^2(s)-f_0^2) G_i G_j  \partial_s z \, ds = \delta_{ij}
\end{equation}
and
\begin{equation}
\label{omega_const_norm_stretched}
 \frac{1}{D} \int_{s(-D)}^{s(0)} F_i F_j  \partial_s z \, ds = \delta_{ij}.
\end{equation}
A necessary condition for using stretched coordinates is that the function $s(z)$ must be strictly monotonic.

\subsection{Density coordinates}
\label{sec:density_coord}
For density coordinates $s = -g \bar{\rho}/\rho_0$, equation \ref{evp_stretched} can be written as
\begin{equation}
 N^4  \partial_{ss}G  +  \partial_z\left(N^2\right) \partial_s G - K^2 G = -\frac{N^2 - f_0^2}{gh}G
\end{equation}
where $F = h N^2 \partial_s G$. Note that the derivatives of the density are computed on the $z$ coordinate, then projected onto the $s$ coordinate in the eigenvalue problem. This avoids using inverses of functions that tend towards zero, and therefore has greater numerical stability. While the method does well for the high wavenumber case (figure \ref{InternalModeError}, lower panel), it performs somewhat poorly with a uniform relative error of $O(0.03)$ for all modes in the low wavenumber (upper panel), as shown by the orange line. Evidently, density coordinates cluster points inefficiently in this case.

\begin{figure}[t]
  \centerline{\includegraphics[width=19pc,angle=0]{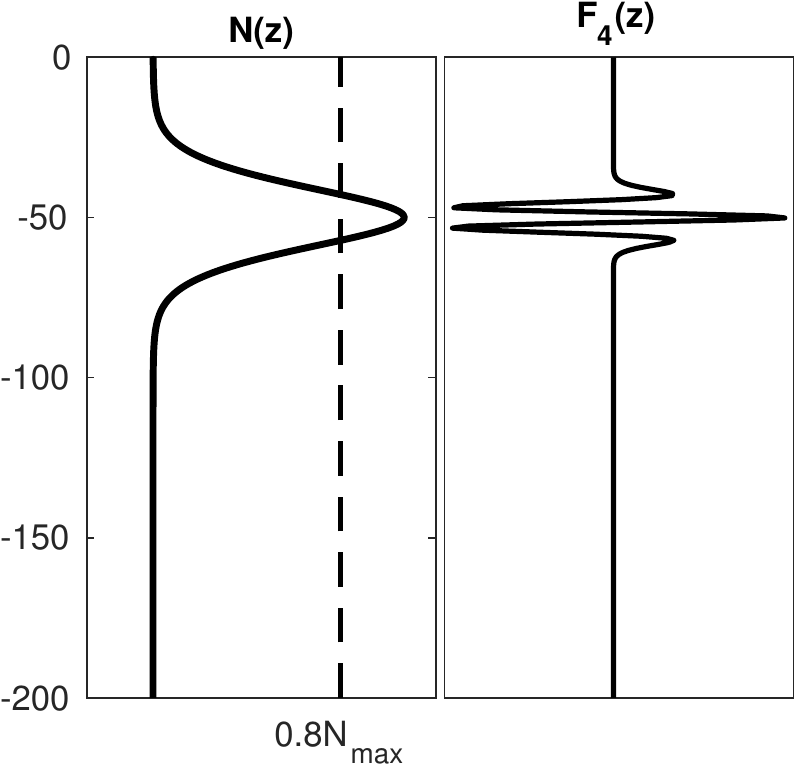}}
  \caption{The left panel shows a stratification profile with pycnocline taken from \citet{cushman2011-book}. The vertical dashed line represents a frequency with two turning points in the pycnocline. The right panel shows the fourth vertical $F$ mode at that frequency.}
  \label{PycnoclineMode}
\end{figure}

\subsection{WKB stretched coordinates}
\label{sec:wkb_coord}
A compromise between  depth ($z$) and  density ($\bar{\rho}$) coordinates is the WKB stretched coordinate, $s = \int_D^z N(z^\prime) \, dz^\prime$. In this case the eigenvalue problem becomes,
\begin{equation}
\label{evp_wkb}
\left( (\partial_z N) \partial_s + N^2 \partial_{ss} \right) G - K^2 G = \frac{f_0^2 - N^2}{gh}G 
\end{equation}
where $F=h N \partial_s G$.

The purple line in figure \ref{InternalModeError} shows that the vertical modes computed on WKB coordinates have uniform accuracy of $O(10^{-3})$ for $K=0$, outperforming the density coordinate case, and also performing nearly as well as density coordinates in the high wavenumber case.

\subsection{Adaptive grid for $\omega$-constant EVP}
\label{sec:adaptive_grid}

\begin{figure}[t]
  \centerline{\includegraphics[width=19pc,angle=0]{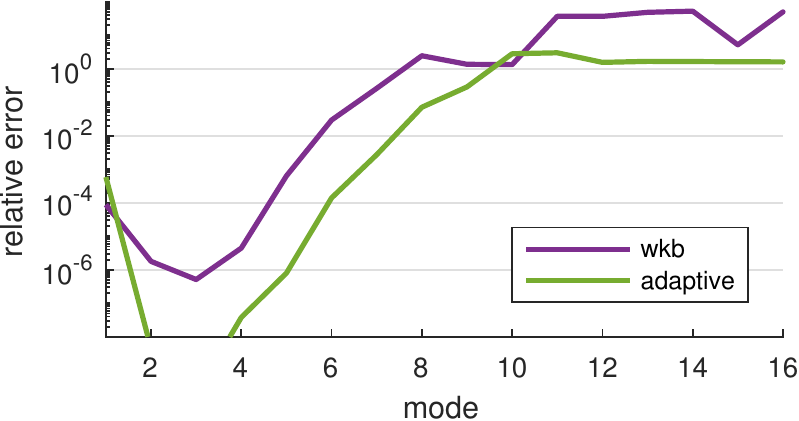}}
  \caption{Relative error as a function of vertical mode number using 64 evenly spaced grid points for the frequency and stratification shown in figure \ref{PycnoclineMode}. Shown is the WKB scaled spectral method (purple) as in figure \ref{InternalModeError}, but also the adaptive grid method (green) that clusters points near the regions of oscillation.Only the first 16 modes are shown.}
  \label{InternalModeErrorFrequency}
\end{figure}

Solving the \ref{vertical-eigenvalue-G-with-omega} suffers from the additional challenge that as the frequency increases and the distance between turning points decreases, the grid spacing necessary to capture the mode structure becomes ever smaller. As noted in the introduction, this issue arises when considering internal waves near a pycnocline. An example stratification profile and vertical mode found at a frequency approaching the maximum frequency in the pycnocline is shown in figure \ref{PycnoclineMode}. The relative error as a function of mode for this example is shown in figure \ref{InternalModeErrorFrequency}, from which it is clear that even the WKB stretched coordinate that performed so well for the \ref{vertical-eigenvalue-G-with-K}, does relatively poorly in this scenario. Sturm-Liouville theory tells us that the $n$-th $F$ mode will have $n$ zero crossings in the oscillatory region where $N^2(z)>\omega^2$ \citep{arfken1970-book}. Thus, in order to resolve these oscillations, one would require at least $2 n$ optimally placed points in that region, as well as additional points to capture the variance in the decay regions. Simply increasing resolution of the Chebyshev grid cannot efficiently solve the problem, as grid points will continue to be poorly placed. 

\begin{figure}[t]
  \centerline{\includegraphics[width=19pc,angle=0]{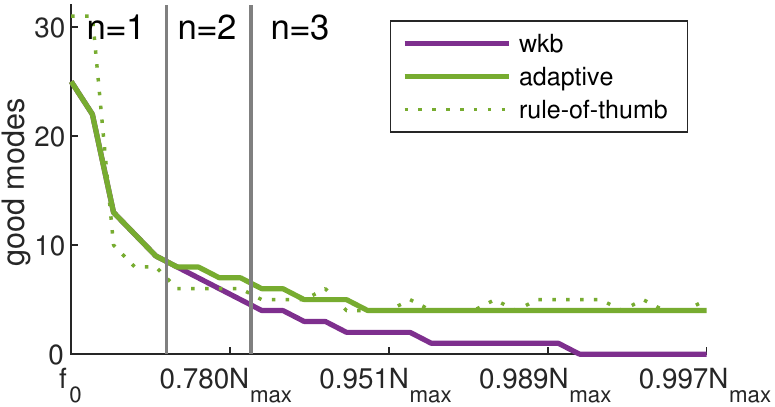}}
  \caption{The number of usable modes (error $< 10^{-2}$) versus frequency for the stratification in figure \ref{PycnoclineMode} using 64 points. The adaptive grid method and standard WKB method are shown in green and purple, respectively. The dashed green line is the rule-of-thumb number of good modes by the adaptive grid method. The two vertical gray lines separate the regions where the adaptive algorithm used 1, 2 or 3 coupled equations. }
  \label{GoodModesVsFrequency}
\end{figure}

To resolve this issue we devise an ad hoc method for clustering points in regions of interest. Our approach is to partition the domain into regions where the modes are hypothesized to be nonzero, formulate the EVP for each region (using WKB stretched coordinates), then couple the equations at the region boundaries. This enables us to assign most of the points to the regions where the solution is assumed to be nonzero, and allocate a few remaining points in the other regions. A comparison of this adaptive method and the standard WKB stretched coordinates is shown in figure \ref{InternalModeErrorFrequency}, where the adaptive method is able to capture a few more usable modes than the standard WKB stretched coordinate method with one EVP. The value of this method becomes more pronounced as the maximum frequency is reached. The number of usable modes (error $< 10^{-2}$) drops to zero as the maximum buoyancy frequency is approached when using the single EVP, as shown in figure \ref{GoodModesVsFrequency}. However, using the adaptive grid algorithm, we are able to guarantee a minimum number of usable modes as points cluster around the turning frequencies. 

The equation boundaries are established by using the WKB approximated solution to identify the regions where the modes are expected to be nonzero. Specifically, the equation boundaries are the points where WKB solution (\ref{Fwkb}) decays to $10^{-5}$ of its value from the turning point. This is an adjustable tolerance, chosen to be small enough that only a few points are needed in the to capture the nearly zero function, but large enough that the nonzero regions aren't unnecessarily large. The gray vertical lines in figure \ref{GoodModesVsFrequency} show the number of coupled equations being used to solve the EVP. At the lowest frequencies only one equation is used and the method is identical to the WKB stretched coordinate method described in section \ref{sec:wkb_coord}. As the frequency increases, the algorithm eventually separates into two coupled equations: one for the top boundary and pycnocline, and another for the deep region where no mode variance is expected (refer to figure \ref{PycnoclineMode}). At high enough frequency the region above the pycnocline is decoupled as well, and three coupled EVP problems are solved.

The EVPs are coupled by requiring that the function and its first derivative are continuous at the equation boundaries, following the procedure described in section 22.3 of \citet{boyd2001-book}. The `eigenvalue rule-of-thumb' as discussed in \citet{boyd2001-book} states that roughly $n/2$ modes will be accurate when using $n+1$ Chebyshev polynomials away from boundary layers or critical levels. Solving the EVP with turning points near the maximum buoyancy certainly does not satisfy this criterion, but the rule-of-thumb can be modified to use half of the modes \emph{with eigenvalues greater than zero}. Although we make no attempt at proving the general validity of this modification, the dashed line in figure \ref{GoodModesVsFrequency} indicates that the rule-of-thumb generally does well at predicting how many modes are good quality.

%
\section{Numerical implementation}
\label{sec:implementation}
%

\begin{figure}
  \centerline{\includegraphics[width=19pc,angle=0]{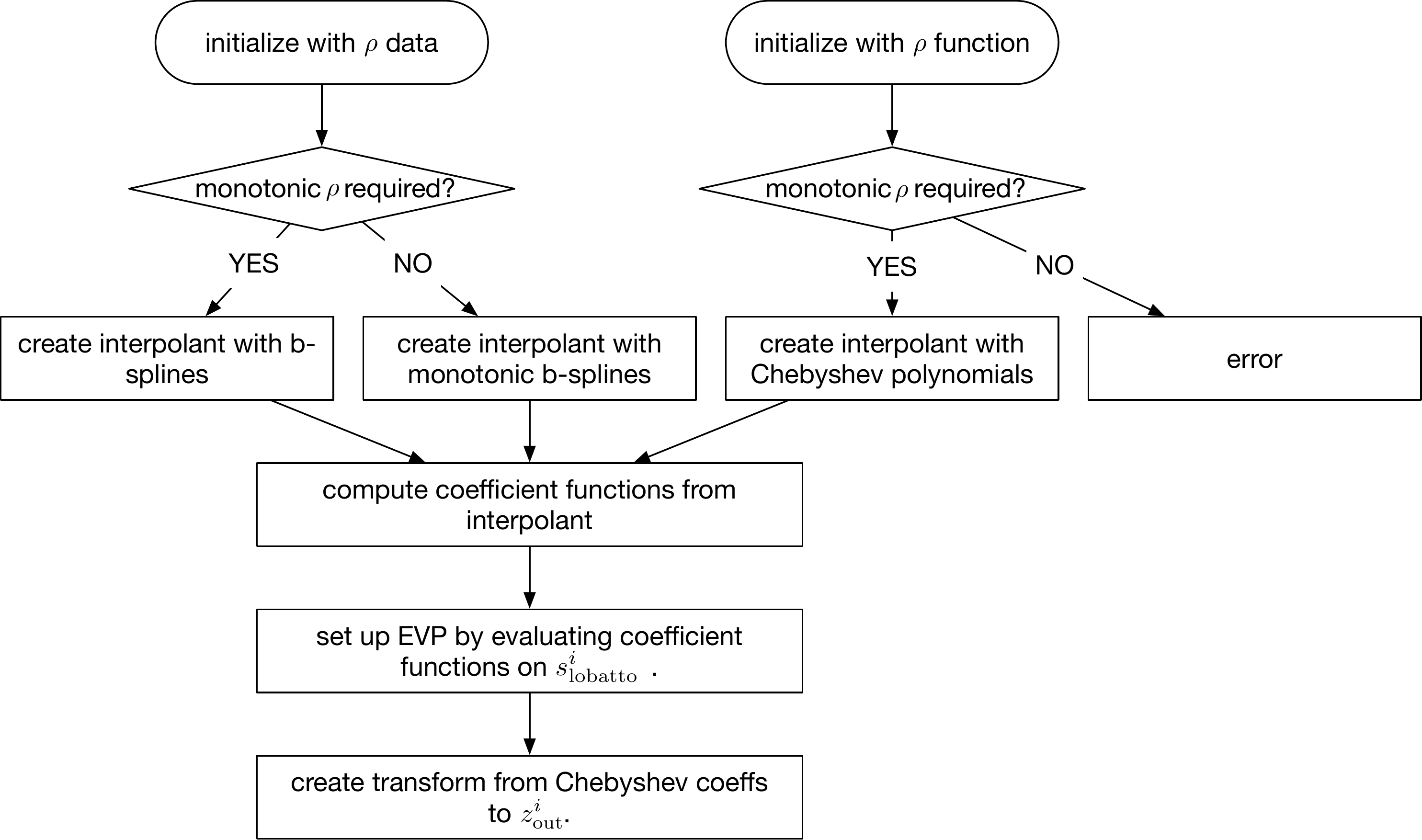}}
  \caption{Algorithm flowchart for the initialization of \texttt{InternalModes} class.}
  \label{InitializationAlgorithm}
\end{figure}

One of the primary products of this paper is the implementation of these methods as classes in Matlab (see \ref{sec:class_hierarchy} for more details). Figure \ref{InitializationAlgorithm} shows the flowchart followed by the initialization algorithm for the \texttt{InternalModes} class, described in this section.

The two methods for initializing the classes are both called using,
\begin{verbatim}
im = InternalModes(rho,z,zOut,latitude);
\end{verbatim}
where the arguments \texttt{rho,z} are either a gridded density field at locations \texttt{z}, or function handle valid in the domain spanned by \texttt{[min(z) max(z)]}. When the function handle is given, the density function is projected onto Chebyshev polynomials. If gridded data is provided, then the density is interpolated using B-splines. The argument \texttt{zOut} specifies the grid on which all output is given, which need not span the full depth.

After initialization, all classes support setting the upper and lower boundary conditions as well as setting the normalization to any of the choices discussed in section \ref{sec:normalization}.

The two primary functions for computing internal modes are
\begin{verbatim}
[F,G,h,omega] = im.ModesAtWavenumber(k);
\end{verbatim}
for the \ref{vertical-eigenvalue-G-with-K} and
\begin{verbatim}
[F,G,h,k] = im.ModesAtFrequency(omega);
\end{verbatim}
for the \ref{vertical-eigenvalue-G-with-omega}.

The implementation of these methods for finite differencing is straightforward---the eigenvalue problem is either solved on the gridded input data as given, or on a grid that matches the output grid if specified as a function. The differentiation matrices are created using the algorithms described in \citet{fornberg1998-siam}.  However, the spectral implementations require additional choices.

The eigenvalue problem being solved is
\begin{equation}
\label{generic_evp}
\left[a(s) \mathsf{T}_{ss} + b(s) \mathsf{T}_s + c(s) \mathsf{T}\right]\vec{v} = \frac{1}{h} d(s) \mathsf{T} \hat{\vec{v}}
\end{equation}
where $s$ is a generic stretched coordinate, $a(s)$, $b(s)$, $c(s)$, and $d(s)$ are referred to here as \emph{coefficient functions}.

The algorithm can be separated into the three parts. First we compute the \emph{coefficient functions} for each eigenvalue problem, e.g., $N^2$ and $\partial_z N$ for equation \ref{evp_wkb}. Second, the eigenvalue problem is solved on the appropriate coordinate with $n_\textrm{evp}$ points. Finally, the resulting modes are normalized and projected onto the output grid with an arbitrary number of points.

\subsection{Initialization with an analytical function}

If the \texttt{InternalModes} class is initialized with a function handle for the density,  it is projected onto Chebyshev polynomials which are then used to compute the coefficient functions and, if necessary, the stretched coordinate.

To project onto Chebyshev polynomials we define a grid with $n_\textrm{grid}$ points using
\begin{equation}\label{lobatto}
z^i_{\textrm{lobatto}} \equiv \frac{z_\textrm{max}-z_\textrm{min}}{2}\left( \cos \left(\frac{i\pi}{n_\textrm{grid}-1} \right) + 1 \right) + z_\textrm{min}
\end{equation}
where $i$ is an integer index ranging from $0$ to $n_\textrm{grid}-1$. We  evaluate the density function on that grid, $\bar{\rho}(z^i_{\textrm{lobatto}})$. The density function is then expanded in a Chebyshev polynomial basis such that,
\begin{equation}
\bar{\rho}(z^i_{\textrm{lobatto}}) = \sum_{k=0}^{n_\textrm{grid}} \hat{\rho}^k T^k(z^i_\textrm{lobatto})
\end{equation}
where $\hat{\rho}^k$ indicates the $k$-th coefficient for Chebyshev polynomial defined on $z$ coordinates. The coefficients for the derivative of the function, denoted $\hat{\rho}_z^k$, are then computed using a recursion formula,
\begin{equation}
\label{cheb_deriv_formula}
c_k \hat{\rho}_z^k = \hat{\rho}_z^{k+2} + 2(k+1)\hat{\rho}^{k+1}
\end{equation}
where $c_k=2$ for $k=0$, and $c_k=1$ otherwise. Because a Gauss-Lobatto grid was used for the $z$-coordinate, the Chebyshev transformation is performed with a rescaled fast Fourier transformation in $O(n_\textrm{grid} \log n_\textrm{grid})$ operations. The differentiation requires only $O(n_\textrm{grid})$ operations, which means that all of the coefficient functions for the eigenvalue problem can be computed on a relatively fine grid. For example, $n_\textrm{grid} = O(2^{14})$ takes a fraction of a second on commodity hardware from 2015.

The stretched coordinates implemented here are either $s=z$, $s = -g\bar{\rho}/\rho_0$, or $s = \int_D^z N(z^\prime) \, dz^\prime$, where the latter two cases require density to be strictly monotonic. For those two cases if $\bar{\rho}_z \geq 0$ anywhere in the domain, then an error is thrown. For the WKB coordinate, $s = \int_D^z N(z^\prime) \, dz^\prime$, the integral is computed spectrally using equation \ref{cheb_deriv_formula}. 

\subsection{Initialization with gridded data}

In the more typical scenario where a user initializes the \texttt{InternalModes} class with gridded data from observations or a numerical model, B-splines are used to interpolate the data and compute the coefficient functions. The primary advantage to using B-splines in this scenario is that B-splines can be created for arbitrary grids without suffering from Runge's phenomena at lower orders. We fit the data to 6th order interpolating spline (with 5 nonzero derivatives) using the methodology and numerical implementation described in \citet{early2019-arxiv}.

If the method requires that density remain monotonic (e.g. for WKB and density coordinates) , then the B-spline fits are constrained to be monotonic following \citet{pya2015-sc}. If the data are not monotonic, then this implicitly smooths to find the nearest monotonic fit in a least-squares sense.

Computing the stretched coordinate and the coefficient functions from the spline interpolant requires derivatives and integrals of the B-splines, which are relatively straightforward to compute because they are just piecewise polynomials \citep{deboor1978-book}. The WKB method requires computing the square root of the B-spline interpolant. The approach taken here is to build a new interpolating spline of the same order that interpolates between the square root of the data points.

\subsection{Eigenvalue problem}

All three Chebyshev methods solve their respective eigenvalue problems on a Gauss-Lobatto grid of their respective coordinate, i.e., $s=z$, $s = -g\bar{\rho}/\rho_0$, or $s = \int_D^z N(z^\prime) \, dz^\prime$. The Gauss-Lobatto grid in $s$ is defined in the usual way as,
\begin{equation}
s^i_{\textrm{lobatto}} \equiv \frac{s_\textrm{max} - s_\textrm{min}}{2}\left( \cos \left(\frac{i\pi}{n_\textrm{evp}-1} \right) + 1 \right) + s_\textrm{min}
\end{equation}
where the number of points $n_\textrm{evp}$ reflects the size of the eigenvalue problem, and is therefore also an absolute upper bound to the number of modes that may be computed.

Once the Gauss-Lobatto grid $s^i_{\textrm{lobatto}}$ is created, the corresponding value $z(s^i_{\textrm{lobatto}})$ is computed using the bisection method as implemented in \citet{chebfun2014-book}. The method is set to terminate with a relative error of $O(10^{-12})$. 

The coefficient functions in equation \ref{generic_evp} are now simply evaluated onto the $s^i_{\textrm{lobatto}}$ grid using the interpolant (either a B-spline or Chebyshev). The Chebyshev polynomials and their derivatives ($T$,$T_s$, and $T_{ss}$) are computed using the standard recursion formulas, and then multiplied by the coefficient functions to create eigenvalue matrices. The boundary conditions are implemented by replacing the first and last rows of the matrices $\mathsf{A}$ and $\mathsf{B}$ in \eqref{EVPforAB}.

The eigenvalue problem is then solved using the standard generalized eigenvalue problem solver. This is typically the rate limiting step in the process, taking $O(n_\textrm{evp}^3)$ operations. The resulting eigenvectors now contain coefficients to the Chebyshev polynomials \emph{defined on the stretched coordinate.}

\subsection{Adaptive grid}

The adaptive grid is created by locating regions in the domain where the solution is expected to be small, and then allocating fewer points (and therefore fewer Chebyshev polynomials) to those regions. After identifying the turning points $z_T$ where $N^2(z_T)=\omega^2$, the WKB solution \ref{Fwkb} is used to identify the equation boundaries $z_{\textrm{bnd}}$, the points where $F^{\textrm{WKB}}_j(z_{\textrm{bnd}},\omega)/F^{\textrm{WKB}}_j(z_{T},\omega)=10^{-5}$. The WKB solution is assumed to work locally in the stratification, and is therefore applied at the turning point, $z_T$, in the direction of decaying variance. The eigenvalue \ref{wkb_eigenvalue} is assumed to be set globally by integration over all oscillatory regions. If $m$ equation boundaries $z_{\textrm{bnd}}$ are identified, they delineate $m+1$ regions: the `decay' regions, where the solution is anticipated to be small and governed by the decaying exponential, and `oscillatory' regions where the solution is expected to dominate and include the oscillatory solution. These decaying and oscillating regions are necessarily alternating.

The $m+1$ regions are coupled using the same technique described in section 22.3 of \citet{boyd2001-book} by requiring continuity across boundaries for $G$ and $\partial_z G$. The key benefit to this algorithm is that the decay regions are allocated as few as 6 points each, while oscillatory regions are apportioned the remaining points relative to their WKB length, $L^m = \int N(z^\prime) dz^\prime$. While we find that 6 points appears to be sufficient for the decay regions, in practice, we do in fact apportion 1/16th of the total points evenly between the decay regions as a hedge that this ad-hoc method will fail for some unforeseen cases.

The adaptive grid algorithms use low-order interpolation to identify $z_T$ and $z_{\textrm{bnd}}$ because high accuracy is not required and this keeps the number of computations  $O(n_\textrm{grid})$.

\subsection{Normalization}

The final step is to normalize the resulting eigenvectors and project them onto $z_\textrm{out}$. Normalization using the two integral conditions can be performed exactly invoking the fact that the integral of each Chebyshev polynomial $T^k(z)$ is exactly known,
\begin{equation}
w^k \equiv \int_{-1}^{1} T^k(z) \, dz = \begin{cases}
\frac{(-1)^k + 1}{1-k^2} & k \neq 1 \\
0 & k=1
\end{cases}.
\end{equation}
We have defined $w^k$ such that it can be summed with a vector of Chebyshev coefficients to produce the integral. In other words, if $\hat{v}^k$ is a Chebyshev coefficient vector, then $I = \sum \hat{v}^k w^k$ is the definite integral. The integrands in equations \ref{k_const_norm_stretched} and \ref{omega_const_norm_stretched} are computed pseudospectrally before integrating (by transforming to the spatial domain, multiplying, then transforming back Chebyshev coefficients).

Computing the max U and and max W norm is more problematic because the function extrema do not necessarily lie on the grid points. For the implementation, we simply take the maximum at the resolved grid points, but if higher accuracy is required, one could locate the extrema using the methods in \citet{boyd2014-book}.

Finally, the normalized eigenmodes are projected onto the output grid using the slow Chebyshev transforms,
\begin{equation}
\label{output_projection}
v(z^i_{\textrm{out}}) = \sum_{k=0}^{n_\textrm{evp}} \hat{v}^k T^k(s(z^i_\textrm{out})).
\end{equation}
If a large number of output points are requested, this operation could dominate the total computation time.

\begin{figure}
  \centerline{\includegraphics[width=19pc,angle=0]{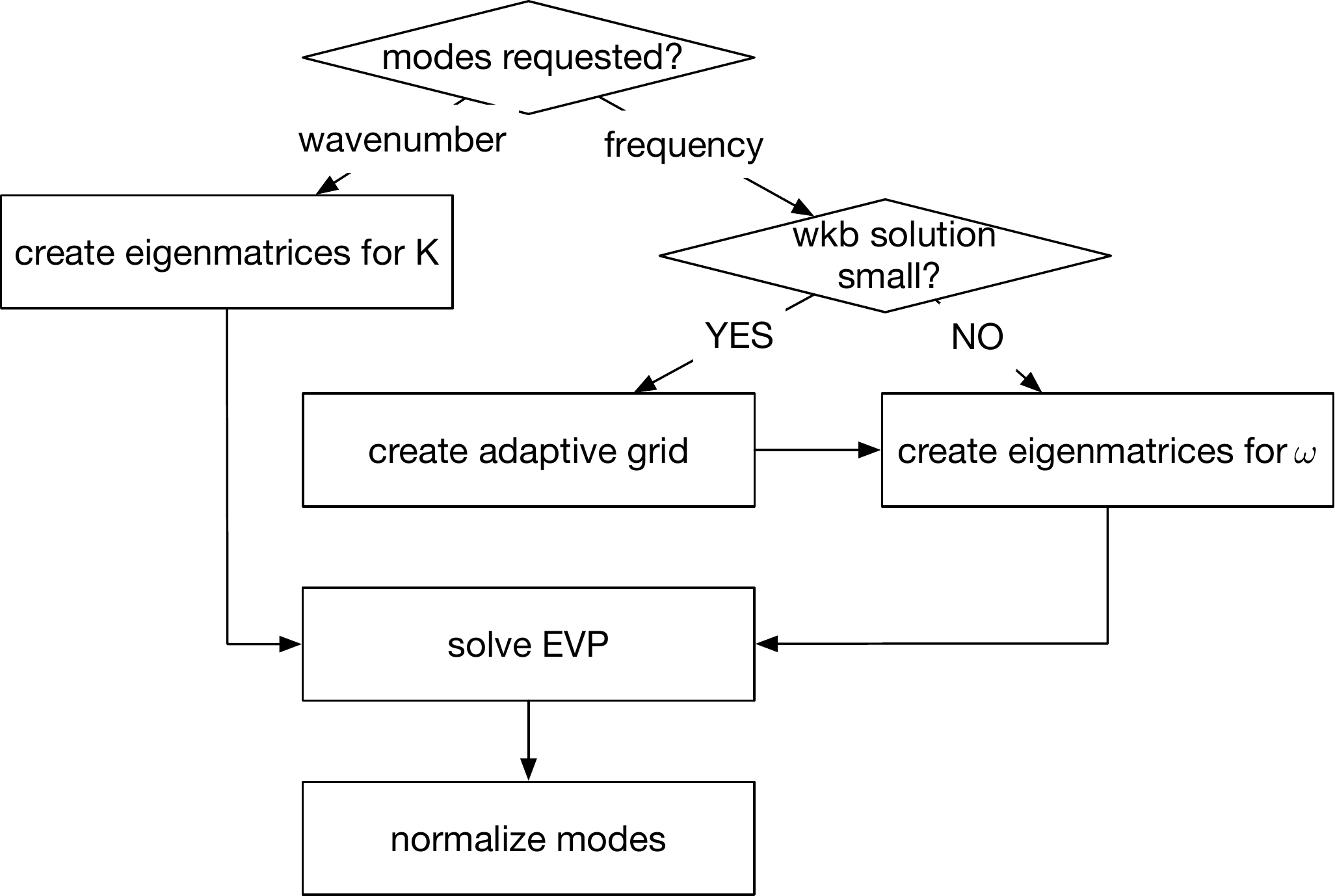}}
  \caption{Algorithm flowchart for the mode computation.}
  \label{ModesAlgorithm}
\end{figure}

The algorithm flowchart for the mode computation is shown in figure \ref{ModesAlgorithm}

\subsection{SQG modes}

The two functions for computing the SQG modes are
\begin{verbatim}
psi = im.SurfaceModesAtWavenumber(k);
\end{verbatim}
and
\begin{verbatim}
psi = im.BottomModesAtWavenumber(k);
\end{verbatim}
where \texttt{k} is an array of wavenumbers.

The SQG modes are found from equation \ref{sqg_cheb_eqn} using a linear solver after replacing the top and bottom points of the matrix with the boundary conditions. As noted in \citet{tulloch2009-jas-note}, the SQG modes require a high density of grid points near the boundaries, a task well suited to the Gauss-Lobatto grid in equation \ref{lobatto}. The number of Chebyshev polynomials is chosen so that the Gauss-Lobatto grid captures at least 10 points over the e-folding scale. The e-fold scale for constant stratification (see appendix \ref{sec:sqg_constant_strat}) is $\Delta z_{\textrm{efold}} = \frac{f_0}{K N_0}$ and the distance between the first two points in a Lobatto grid, equation \ref{lobatto}, is
\begin{equation}
\Delta z_{\textrm{boundary}} = \frac{D}{4} \left( \frac{\pi}{(n_{\textrm{grid}}-1)} \right)^2.
\end{equation}
where we've defined $D=z_\textrm{max}-z_\textrm{min}$ as the depth of the domain. Setting $\Delta z_{\textrm{boundary}} = \frac{1}{10} \Delta z_{\textrm{efold}}$ we find that we need
\begin{equation}
n_{\textrm{grid}} = 1+\frac{\pi}{2} \sqrt{ \frac{10 D K N_0}{ f_0} }
\end{equation}
points (and therefore also $n_{\textrm{grid}}$ polynomials) to sufficiently capture the SQG mode. The resulting SQG modes are projected onto the output grid using equation \ref{output_projection}.

\subsection{Unit testing}

In order to ensure that each of the algorithm implementations is correct, the numerically generated eigenmodes and eigenvalues are compared against the analytical solutions for constant stratification and exponential stratification shown in \ref{sec:analytical_im_solutions}. The comparison is performed across a range of wavenumbers and frequencies for both surface boundary conditions and all four norms.

The computed SQG modes are also compared against analytical solutions for constant and  exponential stratifications, shown in \ref{sec:analytical_sqg_solutions}. 

%
\section{Other sources of error}
\label{sec:error}
%

In the introduction we noted five sources of error that contribute to the total error when computing the forward or inverse transformation, but this manuscript has primarily focused on one source of error: the numerics of accurately representing the vertical modes in the EVP. We now discuss these other sources of error and how they are dealt with in the numerical implementation.

\subsection{Aliasing error}
\label{sec:aliasing_error}

When performing a forward transformation, where a given dynamical field is projected onto the vertical modes, the data grid will determine how many modes are resolvable. As with a Fourier transformation, higher frequency modes alias into the lower frequency modes. Unlike the Fourier transformation, however, the optimal grid for performing a transformation is not an evenly spaced grid, but depends on the stratification profile, and therefore the eigenvalue problem being solved. Here we show that there is a relatively easy method for determining the number of resolvable modes for a given grid using the condition number of the resulting matrix.

To show the effect of different grid choices on the forward transformation, we use the analytical solution of the vertical modes in exponential stratification in combination with an imposed spectrum to generate stochastic isopycnal displacement profiles typical of the world oceans. In particular, we use the Garrett-Munk spectrum,
\begin{equation}
 H(j)= \frac{H_0}{(j + j_\ast)^p}
\end{equation}
where $j_\ast$ is the roll-off mode, usually set to 3 but possibly as high as 20, $p$ is the slope which is usually very nearly $5/2$, and $H_0$ normalizes sum over $j=1..\infty$ to unity. For each stochastically generated set of coefficients, $m^j$, a profile $\eta(z) = \sum_{j=1}^N G^j(z) m^j$ is created. The profile is then evaluated on three different grids: $z^i_\textrm{even}$, $z^i_\textrm{lobatto}$, and $z^i_\textrm{quadrature}$ where $z^i_\textrm{quadrature}$ is the grid of Gaussian quadrature points, determined by the roots of a mode one higher than is trying to be used \citep{press1992-book,boyd2001-book}. Using the first $n$ modes, we then attempt to recover the coefficients using least squares---in practice this is Matlab's \texttt{mldivide} (\texttt{\textbackslash}) operator. For example the first $n$ coefficients of the evenly spaced grid are determined by,
\begin{equation}
\tilde{m}^j_\textrm{even} = G^{j}(z^i_\textrm{even}) \backslash \eta(z^i_\textrm{even})
\end{equation}
where $j=1..n$. The root-mean square error (rmse) is defined as the error in the sum of recovered and missing coefficients,
\begin{equation}
\textrm{rmse}^2 \equiv \frac{\sum_{j=1}^n \left(m^j-\tilde{m}^j\right)^2 + \sum_{j=n+1}^N \left(m^j\right)^2}{\sum_{j=1}^N \left(m^j\right)^2}
\end{equation}

\begin{figure}[t]
  \centerline{\includegraphics[width=19pc,angle=0]{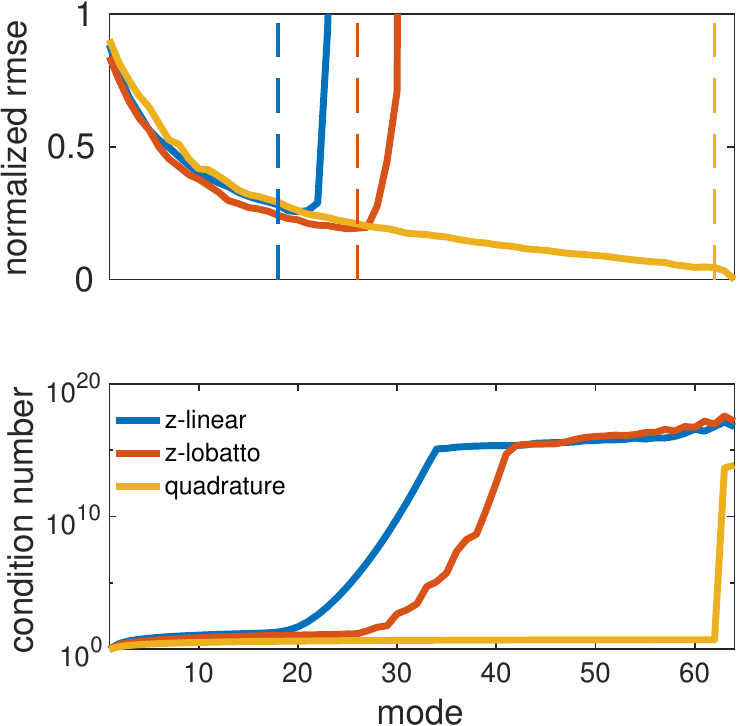}}
  \caption{The top panel shows the root mean square error as a function of total modes used to recover the coefficients with an inverse transformation. Vertical dashed lines are the predicted cutoffs for the different grids, based on the matrix condition number at which the modes are no longer resolvable. The condition number as a function of total modes is shown in the bottom panel.}
  \label{ModeRecoverabilityForDifferentGrids}
\end{figure}

Figure \ref{ModeRecoverabilityForDifferentGrids} shows the result of trying to recover the mode coefficients, $\tilde{m}^j$ using successively more modes for the three different grids. In all three cases the rmse decreases as modes are added until a dramatic increase occurs, correlating with a similarly dramatic increase in the condition number of the matrix. The quadrature grid, defined as the roots of the $G^{N-1}$ mode plus the boundaries, performs best, as expected. Including the boundaries in this definition means that the top and bottom boundary points provide no useful information, and therefore only $N-2$ modes are recoverable for $G$, and $N-1$ for $F$ with a rigid lid. This definition is chosen so that the forward and inverse transform for constant stratification coincides with the discrete sine and cosine transform (and their associated grids).

While the condition number of the matrix is clearly controlled by the grid being used, the choice of norm also affects the condition number. Generally speaking, the \ref{k_const_norm} performs well for transformations with the $G$ modes and the \ref{omega_const_norm} with the $F$ modes. The exception to this is when the free-surface boundary condition is used, the barotropic mode has a substantially different $L^2$ norm and should be rescaled.

\subsection{Mode error}
\label{sec:mode_error}

Here, we test whether or not the truncation error in the vertical modes is a limitation in recovering the coefficients of the spectrum, $m_j$. To this end, we created profiles on a quadrature grid with variance distributed using $H(j)$ for a range of parameters including white noise (large $j_\star$), and very smooth (large $p$), as described above, and recorded how many mode coefficients were recoverable for some error tolerance across all the different numerical methods in this paper. We found that discarding modes with truncation errors exceeding the requested error tolerance of the modes guarantees coefficients recovered with the requested error tolerance. In other words, if one wants relative errors in coefficients of less than $10^{-2}$, one needs modes with errors less than $10^{-2}$. The only exception  is for very steep spectra (e.g., $p=-10$), where the coefficients of the higher modes are indistinguishable from zero. The reverse of this is not true---in fact, including modes with truncation errors exceeding the error tolerance, can often return coefficients within the bounds of the error tolerance. Evidently, the errors in these ordered, orthogonal bases work systematically in our favor. 


\subsection{Interpolation error}

Another source of error may arise from interpolation of the density function. This issue is treated separately from a poorly defined or noisy mean density function and therefore assumes that the data given is gridded and without error. For gridded data, our method uses a relatively low-order B-spline to interpolate between grid points which is then evaluated for the coefficient functions where needed. Does this low-order interpolation method limit the mode recoverability described in the previous section?

To address this question we compute the vertical modes for exponential stratification from an evenly spaced density function with variable number of grid points. These modes are then used to recover the coefficients $\tilde{m}^j$, as above, on a high resolution quadrature grid. We find that with as few as 16 grid points for the density function, we are able to recover 90 mode coefficients with less than 1\% error.


\subsection{Poorly defined or noisy mean density}

It is often the case that the mean density function, $\bar{\rho}(z)$, is not easily defined. Averaging over a mooring time series of density will not necessarily result in a monotonic density function---and averaging often removes the sharp gradients that exist in individual profiles, which may not be desirable. Even output from numerical models can suffer from these same issues, depending on the boundary conditions. Noisy data, where errors in the observed value of $\bar{\rho}(z)$ stem from instrument errors, also effectively constitute a poorly defined mean. One way to frame these issues is to ask how a misspecified mean density affects our ability to infer the vertical spectrum of a given flow.

First we note that all methods, as implemented, will proceed without error for noisy data. However, the most notable difference between methods is that the WKB and density coordinate methods use a density function constrained to the nearest monotonic spline fit, as previously described. It is not clear that this implicit smoothing is necessarily `better' than using the unaltered density function with the z-coordinate method for noisy data. This decision, and how to deal with measurement noise in general, is beyond the scope of this manuscript. Additional smoothing of the density data can be done using many techniques, including the constrained smoothing splines described in \citet{early2019-arxiv}.

In order to test the effects of misspecifying a mean density profile, we  generate density profiles in exponential stratification that follow the GM spectrum, as above, but we attempt to recover the coefficients using vertical modes computed from a noisy mean density profile. The results are consistent with section \ref{sec:mode_error}. For example, as long as the modes from the noisy profile have errors less than $10^{-2}$ relative to the modes generated from the true profile, the recovered coefficients $\tilde{m}_j$ will have errors less than $10^{-2}$ relative to $m_j$. The relative error of the vertical modes increases as a function of mode number until eventually the mode errors and therefore coefficient errors reach $O(1)$. Exactly where mode errors reach $O(1)$ depends on the details of the noise, or misspecification, of the mean density profile.

While accurately recovering the coefficients $m_j$ of the spectrum becomes impossible with a noisy mean density profile, we are able to accurately infer the spectrum from which $m_j$ was generated by either ensemble averaging over additional synthetic profiles, or bin averaging nearby modes. One way to see why this might be true is to note that the coefficient errors never exceed $O(1)$---so although the exact coefficient is incorrect, the magnitude is correct on average. In practice,  using modes from the misspecified mean density profile causes variance that should be associated with mode $j=33$, for example,  to be assigned to the variance of nearby modes.

That the spectrum is recoverable despite a noisy mean is consistent with previous analysis methods. In \citet{polzin2011-rg}, internal wave spectra are found using WKB approximated modes and a WKB stretched grid. It is also important to note that computing the spectrum with a misspecified mean density function still requires an orthogonal set of modes, and therefore fast and accurate mode computation is still helpful.


%
\section{Discussion}
\label{sec:discussion}
%

The spectral methods presented here solve the most relevant forms of the vertical eigenvalue and surface quasigeostrophic mode problems efficiently and accurately. The methods also include an algorithm for computing modes in challenging stratification profiles at high frequencies near turning points. The algorithms are implemented in a publicly available Matlab suite using the class hierarchy described in \ref{sec:class_hierarchy} and the implementations are validated against known analytical solutions, under a wide range of conditions. However, the methods do not always perform well under all conditions.

Poorly resolved features and discontinuities in the density profile will produce the Gibbs phenomenon, where `ringing' occurs in the vicinity of the discontinuity. In one example, we found that a narrow 5-meter-wide pycnocline  in a 5000-meter-deep ocean  produced strong spurious oscillations unless the pycnocline was sufficiently well resolved with enough grid points. In another example, we defined an analytical profile with a discontinuity in $\bar{\rho}_{zz}$ (the highest derivative used in the eigenvalue problem) and this also produced the Gibbs phenomenon. Interestingly, in these cases lower-order finite differencing  produced better modes for the same vertical resolution, because these methods implicitly smooth the derivatives. A logical extension of this work would be to apply the `splitting' algorithms in \texttt{chebfun} \citep{chebfun2014-book} to handle such discontinuities.

Our recommendations for projecting observed or modeled fields onto the modes are as follows,
\begin{itemize}
\item When possible, use a quadrature grid for the fields and modes, e.g. the function \texttt{GaussQuadraturePointsForModesAtWavenumber} in the \texttt{InternalModesSpectral} class computes the Gauss quadrature points for $G$. This is a relatively expensive operation (it involves solving the EVP), but provides near- optimal point placement.
\item In the usual case where there is no freedom to choose grid points, compute the condition number as a function of mode as described in section \ref{sec:aliasing_error}, and limit the number of modes to a low condition number. Alternatively, if computation time is not a limitation, one can perform the least-squares fit of the fields to successively more modes until the coefficients are no longer stable.
\end{itemize}
Many of the errors described in section \ref{sec:error} may still be a concern, but may be quantified with some of the techniques in section \ref{sec:error}, by using the density function and spectrum specific to the problem.



\appendix

%
\section{Internal mode solutions}
\label{sec:analytical_im_solutions}
%

We present analytical solutions for the internal mode eigenvalue problem in three different scenarios: constant stratification, exponential stratification, and the WKB approximated solution for arbitrary stratification. These solutions are used to validate the numerical implementations.

\subsection{Constant stratification}
\label{sec:im_constant_strat}

\begin{table*}
\begin{center}\begin{tabular}{l | l | l | l}\hline
 & trigonometric & linear & hyperbolic \\\hline
$K$-constant & $(-1)^j\sqrt{\sin^2\left(m_jD\right)+\frac{(N_0^2-f_0^2)D}{2g} \left( 1 - \frac{\sin \left( 2 m_j D \right)}{2 m_j D} \right)}$ & $D\sqrt{1+\frac{(N_0^2 - f_0^2)D}{3 g}} $ & $\sqrt{\sinh^2\left(m_0 D\right)+\frac{(N_0^2 - f_0^2)D}{2g}\left(\frac{\sinh(2 m_0 D)}{2m_0 D} - 1 \right) }$ \\
$\omega$-constant & $(-1)^jh_j m_j \sqrt{ \frac{1}{2} + \frac{\sin \left( 2 m_j D \right)}{4 m_j D} }$ & $D$ & $ h_0 m_0 \sqrt{ \frac{\sinh \left( 2 m_0 D \right)}{4 m_0 D} + \frac{1}{2} }$ \\
$U_\textrm{max}$ & $(-1)^jh_j m_j$ & $D$ &  $h_0 m_0 \cosh( m_0 D)$ \\
$W_\textrm{max}$ & $(-1)^j, \sin(m_0 D)\textrm{ for }j=0$ & $D$ & $\sinh(m_0 D)$
\end{tabular}
\caption{ Inverse of the normalization constants ($A^{-1}$) for all combinations of solutions and norms considered here. }
\end{center}
\label{normalizationtable}
\end{table*}

The internal baroclinic modes in constant stratification are given as,
\begin{align} \label{baroclinic_g_mode}
G^{\textrm{const}}_j(z) =& A  \sin \left( m_j ( z + D) \right)\\ \label{baroclinic_f_mode}
F^{\textrm{const}}_j(z) =& A h_j m_j \cos \left( m_j ( z + D) \right).
\end{align}
with eigendepth $h_j$ and vertical wavenumber $m_j$ given by
\begin{equation}
m_j = \frac{j\pi}{D} + \frac{\xi}{D}
\end{equation}
where we have assumed that $w=0$ at the lower boundary $z=-D$. In the case of a rigid lid ($w=0$ at $z=0$), the correction to the vertical wavenumber is $\xi=0$. However, if the linearly approximated free surface boundary condition is used, $h_j G_j(0)=G_j(0)$, then $\xi$ is nonzero. The equations for $\xi$ are transcendental and are therefore solved with a numerical root finding algorithm. The equations for $\xi$ are written in a form conducive for finding the desired root.
\begin{itemize}
\item For fixed wavenumber, $k$, the vertical wavenumber correction $\xi$ is found by solving,
\begin{equation}
\label{baroclinic_fixed_k}
(\xi + j \pi) \left(N_0^2 - f_0^2\right) D \cos(\xi) - g \left( k^2 D^2 + (\xi + j \pi)^2 \right) \sin(\xi) = 0
\end{equation}
near $\xi=0$ and the eigendepth $h_j$ is given by,
\begin{equation}
h_j = \frac{1}{g} \frac{N_0^2 - f_0^2}{k^2+m_j^2}.
\end{equation}
\item For fixed frequency, $\omega$, the vertical wavenumber correction $\xi$ is found by solving,
\begin{equation}
\label{baroclinic_fixed_omega}
D(N_0^2 - \omega^2) - g (\xi + j \pi)\tan(\xi) = 0
\end{equation}
near $\xi=0$ and the eigendepth $h_j$ is given by,
\begin{equation}
h_j = \frac{1}{g} \frac{N_0^2 - \omega^2}{m_j^2}.
\end{equation}
\end{itemize}
The normalization for these modes is given in the first column of table \ref{normalizationtable}.

In the case of the free surface boundary condition, there also exists a barotropic mode ($j=0$), the solution of which changes from trigonometric to hyperbolic at $\omega = N_0$, or $k=k_\ast$ where
\begin{equation}
k_\ast \equiv \sqrt{\frac{N_0^2 - f_0^2}{g D}}.
\end{equation}
In these cases then, the vertical wavenumber reduces to $m_0=\xi/D$.
\begin{itemize}
\item Trigonometric case, $k<k_\ast$ or $\omega < N_0$

The solution is exactly the same as the baroclinic solutions given above in equation \ref{baroclinic_g_mode} and \ref{baroclinic_f_mode}, but now equation \ref{baroclinic_fixed_k} is solved when $j=0$. As a practical matter, it is numerically more stable to solve
\begin{equation}
 \left(N_0^2 - f_0^2\right) D - g \xi^{-1} \left( k^2 D^2 + \xi^2 \right) \tan(\xi) = 0
\end{equation}
for the positive root near $\xi = \sqrt{(N_0^2-f_0^2)D/g - k^2 D^2}$. For fixed frequency, equation \ref{baroclinic_fixed_omega} can be used to find the root near $\xi=\sqrt{D(N_0^2 - \omega^2)/g}$.

\item Linear case, $k=k_\ast$ or $\omega = N_0$

The solution is given by
\begin{align}
G^{\textrm{const}}_0(z) =& A ( z + D)\\
F^{\textrm{const}}_0(z) =& A D
\end{align}
where $h_0=D$.

\item Hyperbolic case, $k>k_\ast$ or $\omega > N_0$

The solution is given by
\begin{align}
G_j(z) =& A  \sinh \left( m_0 ( z + D) \right)\\
F_j(z) =& A h_0 m_0 \cosh \left( m_0 ( z + D) \right).
\end{align}
\begin{itemize}
\item For fixed wavenumber, $k$, the vertical wavenumber correction $\xi$ is found by solving
\begin{equation}
 \left(N_0^2 - f_0^2\right) D - g \xi^{-1} \left( k^2 D^2 - \xi^2 \right) \tanh(\xi) = 0
\end{equation}
for the root near $\xi = \sqrt{k^2 D^2 - (N_0^2-f_0^2)D/g}$ and the eigendepth $h_0$ is given by,
\begin{equation}
h_0 = \frac{1}{g} \frac{N_0^2 - f_0^2}{k^2 + m_0^2}.
\end{equation}
\item For fixed frequency, $\omega$, the vertical wavenumber correction $\xi$ is found by solving
\begin{equation}
 \left(\omega^2 - N_0^2\right) D - g \xi \tanh(\xi) = 0
\end{equation}
for the root near $\xi = \sqrt{D(\omega^2 - N_0^2)/g}$ and the eigendepth $h_0$ is given by,
\begin{equation}
h_0 = \frac{1}{g} \frac{\omega^2 - N_0^2}{m_0^2}.
\end{equation}
\end{itemize}

\end{itemize}

\subsection{Exponential stratification}
\label{sec:im_exponential_strat}

Exponential stratification serves as the $O(1)$ representation of the ocean stratification away from the poles and continental boundaries. As formulated and first solved in \citet{garrett1972-gfd}, we take the stratification to be $N^2 = N_0^2 e^{2z/b}$ where $N_0$ is buoyancy frequency and $b$ is the thermocline depth (with canonical values of 3 cycles per hour and 1300 meters).

Letting the stretched coordinate $s = N_0 e^{z/b}$ as in section \ref{sec:stretched_coord}, the \ref{vertical-eigenvalue-G-with-omega} becomes,
\begin{align}
s^2 G_{ss} + s G_s + \frac{b^2}{c^2} \left( s^2 - \omega^2 \right)G = 0
\end{align}
which has solution
\begin{equation}
G^\textrm{exp}_j (z) = J_\nu \left( \frac{b N_0}{c_j}e^{z/b} \right) - \alpha_j Y_\nu \left( \frac{b N_0}{c_j}e^{z/b} \right)
\end{equation}
where
\begin{equation}
\alpha_j \equiv J_\nu \left( \frac{b N_0}{c_j}e^{-D/b} \right) / Y_\nu \left( \frac{b N_0}{c_j}e^{-D/b} \right)
\end{equation}
is chosen to satisfy the lower boundary condition $G(-D)=0$. The Bessel function $Y_\nu(s)$ has a singularity at $s=0$, so for many choices of $\omega$ and $c_j$, $\alpha_j \ll 1$ and the second term needs to be neglected for stable numerical evaluation. The order of the Bessel function is set by the frequency $\nu=\frac{b \omega}{c}$, or, using the dispersion relation, wavenumber $\nu = \sqrt{ \frac{b^2 f_0^2}{c^2} + b^2 k^2}$. The $F$ modes are found by taking the derivative,
\begin{equation}
\begin{split}
F^\textrm{exp}_j(z) = \frac{N_0}{2g}e^{z/b} \Bigg[  J_{\nu-1} \left( \frac{b N_0}{c_j}e^{z/b} \right) - J_{\nu+1} \left( \frac{b N_0}{c_j}e^{z/b} \right) \\ + \alpha_j \left(Y_{\nu-1} \left( \frac{b N_0}{c_j}e^{z/b} \right) - Y_{\nu+1} \left( \frac{b N_0}{c_j}e^{z/b} \right) \right) \Bigg] 
\end{split}
\end{equation}

The discretization into modes is a result of applying the second boundary condition, in this case either a rigid lid $G(0)=0$ or a free-surface $G(0) = F(0)$, and then finding the eigenmode speeds $c_i = \sqrt{gh_i}$ that satisfy those conditions. The $c_i$'s are therefore determined by the roots of Bessel functions, for which there is no general closed form solution. The challenge then becomes finding bounds for the roots and writing the equation in a form suitable for a root finding algorithm.

In practice it's easiest to find the inverse of the $c_i$'s, so we let $x=\frac{b N_0}{c}$ and then write the root equation, $f(x)=0$, in terms of parameter function $\nu(x)$ and $s(x)$. Again, stable numerical evaluation requires that we work around the singularity of $Y_\nu$, so if $\nu(x)<s(x) e^{-D/b}$, then we take
\begin{equation}
f_{\nu_\textrm{small}}(x) = A(x) J_{\nu(x)}\left( e^{-D/b} s(x) \right) + B(x) Y_{\nu(x)}\left( e^{-D/b} s(x) \right)
\end{equation}
and when $\nu(x)>s(x) e^{-D/b}$ we take
\begin{equation}
f_{\nu_\textrm{big}}(x) = A(x) J_{\nu(x)}\left( e^{-D/b} s(x) \right) / Y_{\nu(x)}\left( e^{-D/b} s(x) \right) + B(x) 
\end{equation}
where in either case the rigid-lid condition requires
\begin{align}
A(x) =& Y_{\nu(x)}\left( s(x) \right) \\
B(x) =& -J_{\nu(x)}\left( s(x) \right) 
\end{align}
and the free-surface requires,
\begin{align}
A(x) =& Y_{\nu(x)}\left( s(x) \right) - \frac{\alpha}{s(x)} \left( Y_{\nu(x)-1}\left( s(x) \right) - Y_{\nu(x)+1}\left(s(x)\right) \right) \\
B(x) =& -J_{\nu(x)}\left( s(x) \right) + \frac{\alpha}{s(x)} \left( Y_{\nu(x)-1}\left( s(x) \right) - Y_{\nu(x)+1}\left(s(x)\right)  \right).
\end{align}

\begin{itemize}
\item $\omega$-constant solution

The parameter functions are defined as,
\begin{align}
\nu(x) =& \frac{\omega x}{\eta} \\
s(x)=& \frac{N_0 x}{\eta}
\end{align}
where the scaling factor $\eta$ is chosen from the WKB approximated solution \citep{desaubies1973-gfd},
\begin{itemize}
\item $\omega > N_0 e^{-D/b}$

The root equation must be taken to be $f_{\nu_\textrm{big}}(x)$ and,
\begin{equation}
\eta \pi = \sqrt{N_0^2 - \omega^2} - \omega \cos^{-1}\left( \frac{\omega}{N_0} \right)
\end{equation}
(from \citet{desaubies1973-gfd} eqn 2.18) the first $n$ modes are found within $[0.5, n+1]$.
\item otherwise

The root equation must be taken to be $f_{\nu_\textrm{small}}(x)$ and,
\begin{equation}
\begin{split}
\eta \pi = \sqrt{N_0^2 - \omega^2} - \omega \cos^{-1}\left( \frac{\omega}{N_0} \right) \\ - \sqrt{N_0^2 e^{-2D/b} - \omega^2} + \omega \cos^{-1}\left( \frac{\omega}{N_0} e^{D/b} \right)
\end{split}
\end{equation}
(from \citet{desaubies1973-gfd} eqn 2.19) the first $n$ modes are found within $[0.5, n+1]$.

\end{itemize}

\item $K$-constant solution

Unlike the $\omega$-constant solution, the order of the Bessel function changes for each root, and therefore the root equation being used may have to change during evaluation.

The parameter functions are defined as,
\begin{align}
\nu(x) =& \sqrt{\epsilon^2 x^2 + \lambda^2} \\
s(x)=& x
\end{align}
where $\epsilon=\frac{f_0}{N_0}$ and $\lambda = b k$. \citet{desaubies1973-gfd} provides low-frequency (lf) and high-frequency (hf) bounds for the eigenvalues, which we can be written in terms of wavenumber,
\begin{align}
x_{\textrm{lf}}(j) =& \left( j - \frac{1}{4} \right) \pi + \lambda \frac{\pi}{2} \\
x_{\textrm{hf}}(j) =& \lambda \left[ 1 + \frac{1}{2} \left( \frac{3\pi(4j-1)}{\lambda 8 \sqrt{2}} \right)^\frac{2}{3} \right]
\end{align}
We transition at
\begin{equation}
x_\nu = \frac{\lambda}{\sqrt{5^2 e^{-2D/b} - \epsilon^2}}
\end{equation}
To set the search bounds for the root algorithm, we want
\begin{equation}
x_\textrm{lower} = 
\begin{cases}
x_{\textrm{lf}}(1) & \lambda < 2(1-\frac{1}{4})\cdot 10^{-1} \\
x_{\textrm{hf}}(1) & \textrm{otherwise}
\end{cases}
\end{equation}
and,
\begin{equation}
x_\textrm{upper} = 
\begin{cases}
x_{\textrm{lf}}(1.1n) & \lambda < n-\frac{1}{4} \\
x_{\textrm{hf}}(5n) & \textrm{otherwise}
\end{cases}
\end{equation}
so the  first $n$ modes are found within $[x_\textrm{lower}, x_\textrm{upper}]$.
\end{itemize}

Normalization is performed by direct evaluation of the mode functions for the $U_\textrm{max}$ and $W_\textrm{max}$, and by numerical integration for the \ref{k_const_norm} and \ref{omega_const_norm} norms.

%
\subsection{WKB solution}
\label{sec:im_wkb_solution}
%

\citet{desaubies1973-gfd} found the WKB solution for stratification with at most one turning point $z_T$. Simplified and using the notation of this manuscript, the WKB solution with single turning point $z_T$ is,
\begin{align}
\label{Gwkb}
G_j^{\textrm{WKB}}(z,\omega) &= A \sqrt{\pi} \left( \frac{\xi}{\omega^2 - N^2}\right)^{\frac{1}{4}} \operatorname{Ai}(\xi)\\ \label{Fwkb}
F_j^{\textrm{WKB}}(z,\omega) &= A \sqrt{\pi} \left( \frac{\xi}{\omega^2 - N^2}\right)^{\frac{1}{4}} \xi_z \operatorname{Ai}^\prime(\xi)
\end{align}
where $\xi = \operatorname{sgn}(\omega^2-N^2) \left(\frac{3}{2} q \right)^{\frac{2}{3}}$ and we've defined
\begin{equation}
\label{q_eqn}
q(z) = \frac{1}{\sqrt{gh_j}} \left| \int_{z_T}^z \sqrt{|N^2-\omega^2|}\,dz \right|
\end{equation}
so that it goes to zero at the turning point, but is positive everywhere else. The eigenvalue is proportional to the integral of the stratification over the oscillatory region,
\begin{equation}
\label{wkb_eigenvalue}
\sqrt{gh_j} = \frac{1}{\left(j-\frac{1}{4}\right)\pi}\left| \int_{z_T}^0 \sqrt{|N^2-\omega^2|}\,dz \right|
\end{equation}
while normalization coefficient for the \ref{k_const_norm} is
\begin{equation}
A(\omega) = (-1)^j \sqrt{2 g} \left[ \int_{z_T}^0 \frac{N^2 - f_0^2}{\sqrt{N^2 - \omega^2} } dz \right]^{-\frac{1}{2}}.
\end{equation}
It is useful to note that away from the turning point, the Airy function $\operatorname{Ai}(\xi)$ can be approximated as sinusoidal above the turning point and exponentially decaying below the turning point,
\begin{multline}
G_j^{\textrm{WKB}}(z,\omega) = \\
\frac{1}{\sqrt{2}} \frac{A(\omega)}{\sqrt[4]{|N^2 - \omega^2|}} \cdot
\begin{cases}
\frac{1}{\sqrt{2}} e^{-q}, & \text{for } z>z_T\\
\sin q + \cos q, & \text{for } z<z_T
\end{cases}
\end{multline}

In the case of no turning point ($\omega < N(z), \forall z \in [-D,0]$) the WKB solution is
\begin{align}
G_j^{\textrm{WKB}}(z,\omega) =& \frac{A(\omega)}{\sqrt[4]{N^2 - \omega^2}} \sin q \\
F_j^{\textrm{WKB}}(z,\omega) =& -\frac{A(\omega) N^2_z h_j}{4(N^2 - \omega^2)^{5/4}} \sin q \\
&+ A(\omega) \sqrt{\frac{h_j}{g}} \sqrt[4]{N^2 - \omega^2} \cos q
\end{align}
where we've taken $z_T = -D$ in \ref{q_eqn}. The eigenvalue is now
\begin{equation}
\sqrt{gh_j} = \frac{1}{j \pi}\left| \int_{-D}^0 \sqrt{|N^2-\omega^2|}\,dz \right|
\end{equation}
Note that this solution does reduce to the exact solution for constant stratification.

%
\section{SQG mode solutions}
\label{sec:analytical_sqg_solutions}
%

The SQG modes are computed by taking the Fourier transform of $\psi$ in equation \ref{sqg_eqn} and scaling by the Fourier transformed boundary condition $\hat{\rho}(k,l,z=0)$ such that $\psi(x,y,z) = \int \int \left(-\frac{g}{\rho_0} \hat{\rho} \bigr\rvert_{z=0}\right)\phi(k,l,z) e^{i(kx+ly)} \, dk \, dl$. Equation \ref{sqg_eqn} then becomes,
\begin{equation}
\label{sqg_wavenumber_eqn}
0 = -K^2 \phi + \frac{\partial}{\partial z} \left( \frac{f_0^2}{N^2} \frac{\partial \phi }{\partial z} \right)
\end{equation}
with surface boundary condition $f_0 \partial_z \phi = 1$ and bottom boundary boundary condition $\partial_z \phi = 0$ (and vice-versa for the bottom boundary modes).

To solve this problem numerically, we project $\phi$ onto a Chebyshev basis where $\vec{v} = \mathsf{T} \vec{\hat{v}}$ is the vector representation of $\phi$.  The equation for the SQG modes becomes
\begin{equation}
\label{sqg_cheb_eqn}
\left[ N^2 \partial_{zz}\mathsf{T}  - \partial_z ( N^2 ) \partial_z \mathsf{T}- \frac{K^2 N^4}{f_0^2} \mathsf{T} \right] \vec{\hat{v}} = 0
\end{equation}
in the interior with boundary conditions $f_0 \partial_z \mathsf{T} \vec{\hat{v}} = 1$ and $f_0 \partial_z \mathsf{T} \vec{\hat{v}} = 0$ for the surface mode.

The numerical implementation of equation \ref{sqg_cheb_eqn} is validated against the known analytical solutions.

\subsection{Constant stratification}
\label{sec:sqg_constant_strat}

The surface quasigeostrophic modes for constant stratification can be found in \citep{tulloch2009-jas}. For the upper boundary this is,
\begin{equation}
f_0  \phi_{\textrm{sur}}^{\textrm{const}}(z) = \frac{1}{\lambda} \frac{\cosh\left( \lambda (z+D) \right) }{\sinh\left( \lambda D \right)}
\end{equation}
and the lower boundary,
\begin{equation}
f_0  \phi_{\textrm{bot}}^{\textrm{const}}(z) = -\frac{1}{\lambda} \frac{\cosh\left( \lambda z \right) }{\sinh\left( \lambda D \right)}
\end{equation}
where $\lambda = \frac{KN_0}{f_0}$. These solutions cannot be evaluated numerically for all wavenumbers and depths because the $\sinh$ function may overflow. Instead, we use
\begin{equation}
f_0  \phi_{\textrm{sur}}^{\textrm{const}}(z) = \frac{1}{\lambda} \frac{e^{\lambda z} + e^{-\lambda(z+2D)}}{1 - e^{-2\lambda D}}
\end{equation}
and
\begin{equation}
f_0  \phi_{\textrm{bot}}^{\textrm{const}}(z) = -\frac{1}{\lambda} \frac{e^{\lambda (z-D)} + e^{-\lambda(z+D)}}{1 - e^{-2\lambda D}}
\end{equation}
for reliable numerical evaluation.

\subsection{Exponential stratification}
\label{sec:sqg_exponential_strat}

The surface SQG modes for exponential stratification are first solved by \citet{lacasce2012-jpo}. Defining the scale $\eta \equiv \frac{N_0 K b}{2 f_0}$, the surface mode is given by
\begin{equation}
\phi_{\textrm{sur}}^{\textrm{exp}}=\frac{e^\frac{z}{b}}{N_0 K} \frac{K_0\left(2\eta e^{-\frac{D}{b}}\right)I_1\left(2\eta e^\frac{z}{b}\right) + I_0\left(2\eta e^{-\frac{D}{b}}\right) K_1\left(2\eta e^\frac{z}{b}\right)}{I_0(2\eta)K_0\left(2\eta e^{-\frac{D}{b}}\right) - K_0(2\eta) I_0\left(2\eta e^{-\frac{D}{b}}\right)}
\end{equation}
while the bottom mode can be shown to be,
\begin{equation}
\phi_{\textrm{bot}}^{\textrm{exp}}=\frac{e^\frac{z+2D}{b}}{N_0 K} \frac{K_0(2\eta)I_1\left(2\eta e^\frac{z}{b}\right) + I_0(2\eta) K_1\left(2\eta e^\frac{z}{b}\right)}{K_0(2\eta)I_0\left(2\eta e^{-\frac{D}{b}}\right) - I_0(2\eta) K_0\left(2\eta e^{-\frac{D}{b}}\right)}.
\end{equation}
Bessel functions $K_0$ and $K_1$ should not be confused with the wavenumber $K$.

\subsection{WKB solution}
\label{sec:wkb_sqg_solution}

The WKB solution for the SQG mode does not appear to have been previously derived. Assuming a solution of the form $\phi = e^{q(z)}$ and inserting this into equation \ref{sqg_eqn}, we have that,
\begin{equation}
q^{\prime\prime} + {q^\prime}^2 - 2 \left( \ln N \right)^\prime q^\prime - N^2 K^2 = 0.
\end{equation}
We then assume a series expansion $q=q_0 + \epsilon q_1 + ...$ combined with the assumption that both $q^{\prime\prime}$ and $N^\prime$ are $O(\epsilon)$. The two lowest order equations are therefore,
\begin{align}
{q_0^\prime}^2 - N^2 K^2 =& 0 & O(1) \\
q_0^{\prime\prime} + 2 q_0^\prime q_1^\prime - 2 \left( \ln N \right)^\prime q_0^\prime &= 0 & O(\epsilon)
\end{align}
which has solution,
\begin{equation}
\phi(z) \approx \sqrt{ \frac{N(z)}{N_0} } \left[ A  e^{\frac{K}{f_0} \int_{-D}^z N(z)} + B e^{-\frac{K}{f_0} \int_{-D}^z N(z)} \right].
\end{equation}
Applying boundary condition $f_0 \phi_z(-D) = 0$ and $f_0 \phi_z(0) = 1$, the surface mode is
\begin{align}
\phi_{\textrm{sur}}^{\textrm{WKB}}(z) \approx& \frac{1}{K N_0}  \sqrt{  \frac{N(z)}{N_0} } \left[\frac{\alpha  e^{s(z)} +   e^{\left( -s(z)  + 2 s(-D)\right)} }{ \alpha \left( b + 1 \right) +  \left(b - 1  \right)e^{2s(-D)} }\right]
\end{align}
where $s(z) \equiv \frac{K}{f_0} \int_{0}^z N(z) \, dz$ in the integral from the surface to depth, while $a=\frac{f_0 N_z(-D)}{2K N^2(-D)}$, $b=\frac{f_0 N_z(0)}{2K N^2_0}$, and $\alpha = \frac{1-a}{1+a}$.

For constant stratification this reduces to the result in appendix \ref{sec:sqg_constant_strat}.

%
\section{Class hierarchy}
\label{sec:class_hierarchy}
%

\begin{figure}[t]
  \centerline{\includegraphics[width=19pc,angle=0]{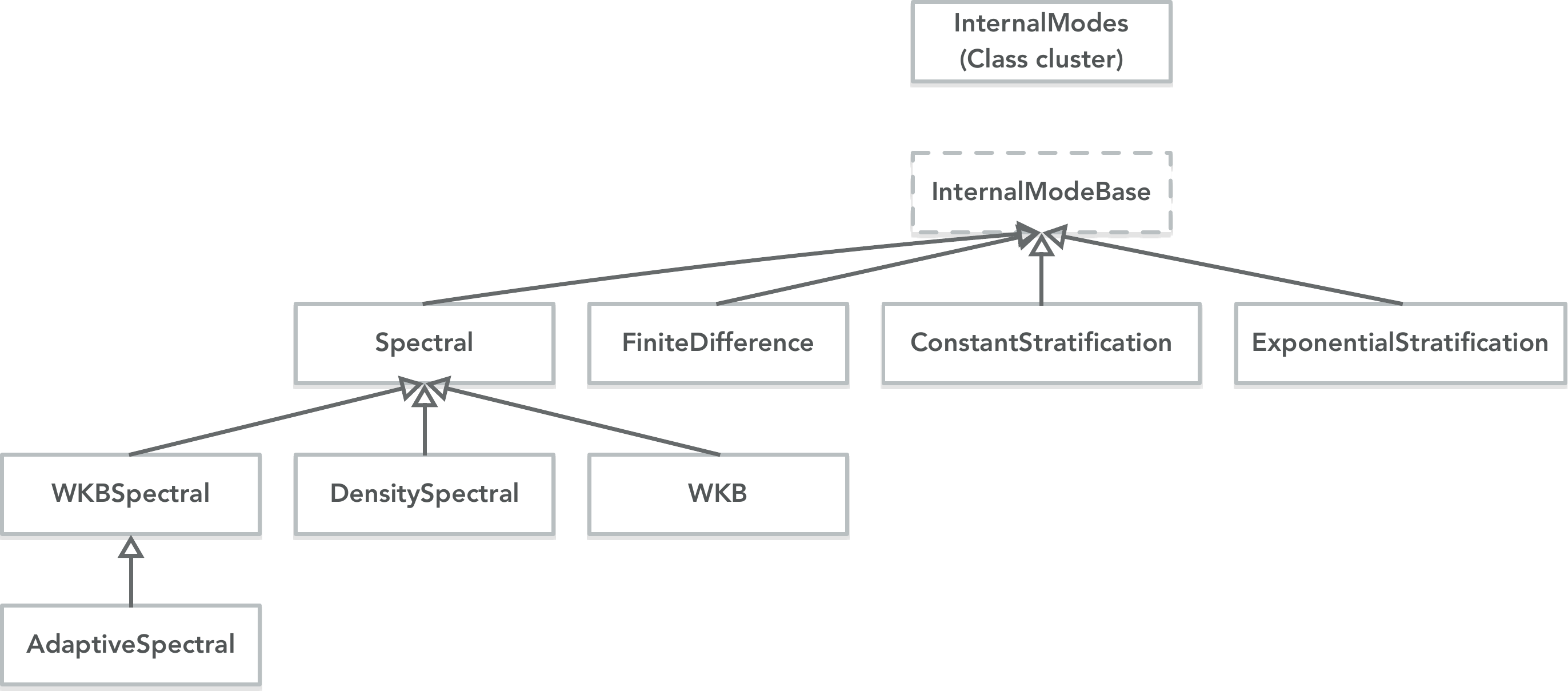}}
  \caption{Class hierarchy for the Matlab implementation of the algorithms in this manuscript. \texttt{InternalModesBase} is an abstract class.}
  \label{ClassHierarchy}
\end{figure}

The algorithms in this manuscript are implemented as classes in Matlab in order to take advantage of class-based inheritance. The class hierarchy is shown in figure \ref{ClassHierarchy}, where \texttt{InternalModeBase} is the abstract superclass which defines the primary interface. A class cluster \texttt{InternalModes} (not part of the hierarchy) is included to provide a single interface from which to initialize all the concrete subclasses.

The primary \texttt{Spectral} class uses depth ($z$) coordinates from section \ref{sec:z_coord} to compute the eigenvalue problem, while the \texttt{DensitySpectral} and \texttt{WKBSpectral} subclasses use the stretched coordinates described in sections \ref{sec:density_coord} and \ref{sec:wkb_coord}, respectively. The \texttt{AdaptiveSpectral} class overrides its superclass when the frequency is high enough, as described in section \ref{sec:adaptive_grid}. The \texttt{ConstantStratification} class implements the analytical solution from \ref{sec:im_constant_strat} and \ref{sec:sqg_constant_strat}, while the \texttt{ExponentialStratification} class implements the analytical solution from \ref{sec:im_exponential_strat} and \ref{sec:sqg_exponential_strat}. The WKB class implements the WKB approximated solution from \ref{sec:im_wkb_solution} and \ref{sec:wkb_sqg_solution} and inherits from the \texttt{Spectral} class in order to use the spectrally computed stratification function, $N^2$. The \texttt{FiniteDifference} class uses finite difference matrices of arbitrary order, on arbitrary grids using the algorithm described in \citet{fornberg1998-siam}.

\section*{Acknowledgments}

All classes available at \texttt{https://github.com/JeffreyEarly/GLOceanKit/}.

Thanks to Jonathan Lilly for his careful comments on an early draft of this manuscript. This work was supported by Office of Navy Research grant N00014-15-1-2465, and National Science Foundation award numbers 1658564 and 1536747.

\bibliographystyle{ametsoc2014}

\end{document}